\documentclass[preprint,12pt]{elsarticle}

\usepackage[T1]{fontenc}
\usepackage{lmodern}
\usepackage[utf8]{inputenc}
\usepackage{booktabs,array}
\usepackage{calc}
\usepackage{graphicx}
\usepackage{amsmath,amssymb}
\usepackage[section]{placeins}
\usepackage{flafter}
\usepackage{url}
\usepackage{xurl}
\usepackage{hyperref}
\hypersetup{hidelinks}

\graphicspath{{04_figures/}}

\makeatletter
\g@addto@macro\UrlBreaks{\do\+}
\makeatother

\journal{Expert Systems with Applications}
\biboptions{sort&compress}

\begin{document}

\begin{frontmatter}

\title{AutoQuant: An Auditable Expert-System Framework for Execution-Constrained Auto-Tuning in Cryptocurrency Perpetual Futures}

\newif\ifblind
\blindfalse

\ifblind
\author{Anonymous author(s)}
\else
\author{Kaihong Deng}
\affiliation{organization={School of Modern Information Industry, Artificial Intelligence Program, Guangzhou College of Commerce},
            city={Guangzhou},
            postcode={510555},
            country={China}}
\ead{202312100202@xs.gcc.edu.cn}
\fi

\begin{abstract}
Backtests of cryptocurrency perpetual futures are sensitive to execution
timing, funding alignment, trading costs, and reuse of evaluation windows
during parameter search. In high-friction markets, attractive results may
therefore reflect hidden implementation choices as much as signal quality.
Using BTC/USDT, ETH/USDT, SOL/USDT, and AVAX/USDT perpetual contracts, this
study examines whether an auditable execution-aware configuration-selection pipeline
can reduce performance overestimation and expose parameter fragility more
clearly than naive one-stage tuning.

This paper proposes AutoQuant, an expert-system-style decision-support
framework for configuration selection. AutoQuant encodes strict execution
timing, funding visibility, cost realism, and feasibility constraints as
explicit rules; combines Bayesian search with two-stage screening across
windows and cost scenarios; and exports deterministic artifacts with
accounting-invariant checks for traceability. The resulting governance
protocol selects and documents configurations within a pre-specified signal
family under strict semantics.

Empirically, fee-only and zero-cost backtests materially inflate apparent
performance relative to fully costed runs with funding and slippage. Two-stage
screening does not guarantee higher returns; in the BTC anchor case and
several replications, it more often surfaces lower-drawdown or less extreme
alternatives under identical strict semantics. Same-budget optimizer
comparison, module and screening-policy ablations, funding-rule diagnostics,
inferential checks, cross-asset replications, and third-party replay checks
position AutoQuant as auditable validation infrastructure for configuration
selection under explicit execution and cost assumptions. The experiments use
small-account simulations under linear costs and exclude market impact and
institutional capacity constraints.

\end{abstract}


\end{frontmatter}

\section{Introduction}

\noindent\textbf{Motivation and background.}

The proliferation of cryptocurrency derivatives has fundamentally
reshaped the digital asset landscape, with perpetual futures contracts
emerging as a dominant instrument for both retail and institutional
participants. Traded 24/7 on global exchanges, instruments such as the
BTC/USDT perpetual exhibit high volatility, complex non-linear dynamics,
and periodic liquidity crises. The market's microstructure introduces
unique challenges for quantitative strategy development, most notably
through its intricate cost structure. Unlike traditional equity markets,
a strategy's profitability is co-determined by a mix of bilateral
trading fees, periodic funding payments that anchor the contract price to
the spot index, execution slippage, and stringent constraints on leverage
and notional position sizes. The confluence of these factors makes the
journey from a research backtest to live trading exceptionally fragile.
From an expert-systems and decision-support perspective, the reliability
of backtests in such markets directly affects how organizations build,
validate, and govern algorithmic trading and risk-control components that
may be embedded in digital-asset products, automated trading platforms,
and DeFi-style protocols.

Within this high-stakes environment, the processes of backtesting and
parameter tuning are prone to significant bias. Common methodological
pitfalls frequently lead to substantial overestimation of performance.
These include \textbf{look-ahead bias}, which arises from the improper use
of future information, and \textbf{unrealistic cost assumptions}, where
critical factors such as funding payments, slippage, and leverage
constraints are ignored or oversimplified. Furthermore, many strategies
rely on manual, \emph{ad hoc} parameter tuning, yielding configurations
that appear profitable in-sample but fail to generalize to
out-of-sample data or to the unforgiving conditions of live markets. This
persistent gap between research and reality motivates the development of
an implementation-oriented framework designed explicitly to bridge the
divide between promising backtests and deployable strategies for
cryptocurrency perpetuals.

\noindent\textbf{Scope and interpretation.}
The empirical scope is deliberately bounded. AutoQuant is positioned as a
configuration-selection and traceability protocol for a pre-specified signal family,
with evidence from four liquid perpetual contracts (BTC/USDT, ETH/USDT,
SOL/USDT, and AVAX/USDT). The cross-asset results serve as portability checks
for the validation protocol rather than broad market validation.

\noindent\textbf{Paper organization.}
The paper proceeds as follows. Section~\ref{sec:methodology} presents the
core methodology of the execution-centric auto-tuning and double-screening
framework. The related literature is then reviewed, followed by the data
and market setting, system architecture and implementation, empirical
evaluation, discussion of implications and limitations, and the final
conclusion.

\noindent\textbf{Reproducibility pointer and timestamp convention.}
All core evidence in this manuscript is generated from deterministic
machine-readable artifacts produced by the pipeline. Appendix~A lists
minimal scripts and commands to regenerate the
reported tables and figures, subject to the data-access constraints
described there. All date ranges are expressed in UTC and refer to the
\emph{bar-open timestamp} used to index the OHLCV series. Under STRICT4H,
signals are computed at bar close and, by design, take effect only from
the subsequent bar.

This paper addresses a narrow but practically important question: under a
strict 4-hour perpetual-futures protocol, does execution-aware,
double-screened configuration selection reduce the performance inflation and
parameter fragility that commonly arise under naive one-stage tuning and
simplified costs? AutoQuant is designed to answer this question in an
auditable way. Rather than treating execution timing, funding visibility,
trading costs, and screening policy as informal implementation details, it
turns them into explicit system rules whose effects can be inspected,
stress-tested, and replayed. Empirically, the study is anchored on BTC/USDT and
extends the same protocol to ETH/USDT, SOL/USDT, and AVAX/USDT as bounded
portability checks.

From an expert-systems and decision-support perspective, this paper therefore positions
AutoQuant as validation and governance infrastructure for
execution-constrained configuration selection in cryptocurrency perpetual
futures. The system encodes strict execution semantics, funding alignment,
cost realism, and feasibility constraints as inspectable policy, and exports
deterministic artifacts and accounting-invariant checks to support traceable
offline-to-live consistency. In high-friction perpetual markets, the central
claim is narrower: credible backtest evidence requires configuration
selection rules that are explicit enough to be inspected and replayed.

\paragraph{Contributions}
The contributions are fourfold. First, this paper formulates
configuration selection for frictional perpetual-futures strategies as an
auditable expert-system-style decision-support problem in which execution
semantics, funding visibility, trading costs, and feasibility constraints are
encoded as explicit system rules rather than treated as informal backtest
choices. Second, it implements a policy-driven inference pipeline that combines
Bayesian search with a two-stage double-screening protocol under strict
\(t\!+\!1\) execution semantics and explicit cost-mis-specification scenarios.
Third, it introduces a traceability layer based on deterministic
machine-readable artifacts, accounting-invariant checks, and offline-to-live
consistency mechanisms, so that configuration selection can be replayed and
traced end-to-end. Fourth, through a BTC/USDT anchor case together with
cross-asset replications on ETH/USDT, SOL/USDT, and AVAX/USDT, it shows how the
framework exposes cost-driven performance inflation, window fragility, and
residual overfitting risk under a unified strict protocol.

\noindent\textbf{Nomenclature and abbreviations.}
Throughout the manuscript, \texttt{STRICT4H} denotes the strict 4-hour
execution protocol used in the case study; \texttt{OHLCV} denotes open, high,
low, close, and volume bar data; \texttt{TPE} denotes the Tree-structured
Parzen Estimator sampler used in Stage~I; \texttt{WFE} denotes walk-forward
evaluation; \texttt{CAGR} denotes compound annual growth rate; \texttt{MaxDD}
denotes maximum drawdown; \texttt{DSR} denotes the deflated-Sharpe-style
diagnostic used in the long-window analysis; and \texttt{PBO} denotes
probability of backtest overfitting.

\noindent\textbf{Limitations of existing practice.}

While numerous open-source frameworks such as Backtrader \citep{backtrader},
Zipline \citep{zipline}, Freqtrade \citep{freqtrade}, and FinRL
\citep{finrlgithub} have democratized algorithmic trading research, they
typically prioritize strategy prototyping over strict realism. In many equity
or low-friction settings this is acceptable if the user configures the system
carefully. The concern here is narrower: in high-friction cryptocurrency perpetual
markets, simplified defaults can yield backtests that look attractive on paper
but do not survive realistic cost and execution modeling. In official
repositories and user guides, critical choices about execution semantics and
cost realism are usually left to the end user. As a result, no-look-ahead
invariants may not be enforced architecturally, cost models are often reduced
to fees with only partial support for funding and leverage constraints, and
parameter search is easily decoupled from realistic risk controls.

These frameworks are not treated as flawed in general. The point is that
without strict guardrails, common usage patterns in perpetual-futures studies
can create a substantial illusion of profitability. Appendix~A illustrates the
implementation burden with both a minimal Backtrader raw-return semantics
check and a stronger Backtrader event-loop replay over the executed STRICT4H
exposure path. These checks require bespoke data plumbing, timestamp
alignment, explicit funding inputs, consistent fee/slippage/leverage
constraints, deterministic export of per-bar return and cost components, and
full-chain accounting invariants. They are therefore useful portability
evidence for adapter-based traceability in generic trading frameworks.

\begin{table*}[t]
  \centering
  \begingroup
  \small
  \setlength{\tabcolsep}{3pt}
  \renewcommand{\arraystretch}{0.95}
  \caption{\textbf{AutoQuant versus common trading research frameworks (feature matrix).} Entries indicate whether a feature is \emph{native}, \emph{configurable}, or absent (\(\times\)). This classification reflects the capabilities described in official documentation and common usage patterns, and is not intended as an exhaustive survey of all possible extensions.}
  \label{tab:framework_matrix}
  \resizebox{\textwidth}{!}{%
  \begin{tabular}{@{}lccccccc@{}}
    \toprule
    Framework & Funding & \shortstack{\(t+1\)\\semantics} & \shortstack{Scenario-grid\\stress} & \shortstack{Bayesian\\tuning} & \shortstack{Full-chain\\invariants} & \shortstack{Guard\\overlay} & \shortstack{Perp-specific\\constraints} \\
    \midrule
    AutoQuant (this paper) & native & native & native & native & native & native & native \\
    Backtrader \citep{backtrader} & configurable & configurable & \(\times\) & configurable & \(\times\) & configurable & configurable \\
    Zipline \citep{zipline} & \(\times\) & configurable & \(\times\) & configurable & \(\times\) & \(\times\) & \(\times\) \\
    Freqtrade \citep{freqtrade} & \(\times\) & configurable & \(\times\) & configurable & \(\times\) & configurable & \(\times\) \\
    FinRL \citep{finrlgithub} & \(\times\) & configurable & \(\times\) & configurable & \(\times\) & configurable & \(\times\) \\
    \bottomrule
  \end{tabular}
  }
  \endgroup
\end{table*}

\noindent\textbf{Why generic research frameworks are not enough in this setting.}
The comparison with common research frameworks is not intended to argue that
such frameworks are unusable. The point is narrower and more operational:
in high-friction perpetual-futures markets, credible evaluation requires that
execution delay, funding visibility, leverage limits, notional caps, and
per-bar cost accounting be treated as enforceable semantics rather than as
optional user discipline. The practical contribution of AutoQuant is therefore
to reduce the implementation gap between a backtest that can be run and one
whose semantics can be checked and replayed.

A more fundamental limitation is that much prior academic and practitioner
work remains \textbf{alpha-centric}. The main emphasis is usually on new
predictive signals or factors, while the validation pipeline is treated as a
secondary concern. Comparatively less attention is paid to the systematic,
automated, and robust generation and validation of the \emph{parameters} that
govern these signals. The machinery that transforms a raw strategy idea into a
vetted, tradable candidate therefore remains under-explored.

\section{\texorpdfstring{Methodology: Execution-Centric Auto-Tuning\\and Double Screening}{Methodology: Execution-Centric Auto-Tuning and Double Screening}}\label{sec:methodology}

This section presents AutoQuant as an execution-aware decision-support
protocol for robust configuration selection. The emphasis is not on inventing
a new signal family, but on making the assumptions that govern evaluation and
selection explicit, enforceable, and reproducible. Concretely, the methodology
combines a bounded Stage~I search, a Stage~II double-screening procedure under
strict semantics, and a deployment-oriented supervision layer that together
turn fragile backtest choices into inspectable system policy.

\paragraph{Expert-system decomposition}
For readers from the expert-systems and applied AI community, AutoQuant should
be understood not as a symbolic reasoner for market prediction, but as a
rule-explicit decision-support framework for robust configuration selection.
Its knowledge base comprises enforceable operational rules: strict
\(t\!+\!1\) execution, no-look-ahead funding alignment, explicit cost and
constraint profiles, and ex ante screening thresholds. Its inference engine is
the search-and-screen pipeline that evaluates candidate configurations under
those rules. Its explanation and governance layer exports deterministic
artifacts, reconciliation checks, and replay-style supervision signals so that
the resulting recommendation can be independently traced and verified. In this
sense, the contribution is methodological and architectural: AutoQuant turns
fragile backtest assumptions into inspectable system policy.

\begin{table}[!htbp]
  \centering
  \footnotesize
  \setlength{\tabcolsep}{3pt}
  \renewcommand{\arraystretch}{1.0}
  \caption{\textbf{ESWA-aligned decision-support mapping of AutoQuant components.}}
  \label{tab:eswa_mapping}
  \begin{tabular}{@{}p{2.8cm}p{3.8cm}p{5.8cm}@{}}
    \toprule
    ESWA component & AutoQuant realization & Decision-support / traceability output \\
    \midrule
    Knowledge base & STRICT4H execution semantics; funding alignment; cost/constraint profiles; screening policies & Declarative rules that prevent look-ahead and enforce feasibility; explicit cost-scenario grid and window semantics. \\
    Inference engine & Stage~I Bayesian optimization (TPE) + Stage~II policy-driven double-screening & Stable configuration recommendation with robustness diagnostics across windows and cost-mis-specification scenarios. \\
    Explanation interface & Deterministic machine-readable exports; accounting invariants; full-chain consistency checks & Traceable artifacts for table/figure regeneration and backtest-to-live consistency checks. \\
    Governance overlay & Guard triggers + invariant monitoring & Run-time supervision signals and reconciliation deltas to detect drift and implementation inconsistencies. \\
    \bottomrule
  \end{tabular}
\end{table}

\begin{figure}[!htbp]
  \centering
  \includegraphics[width=0.92\linewidth]{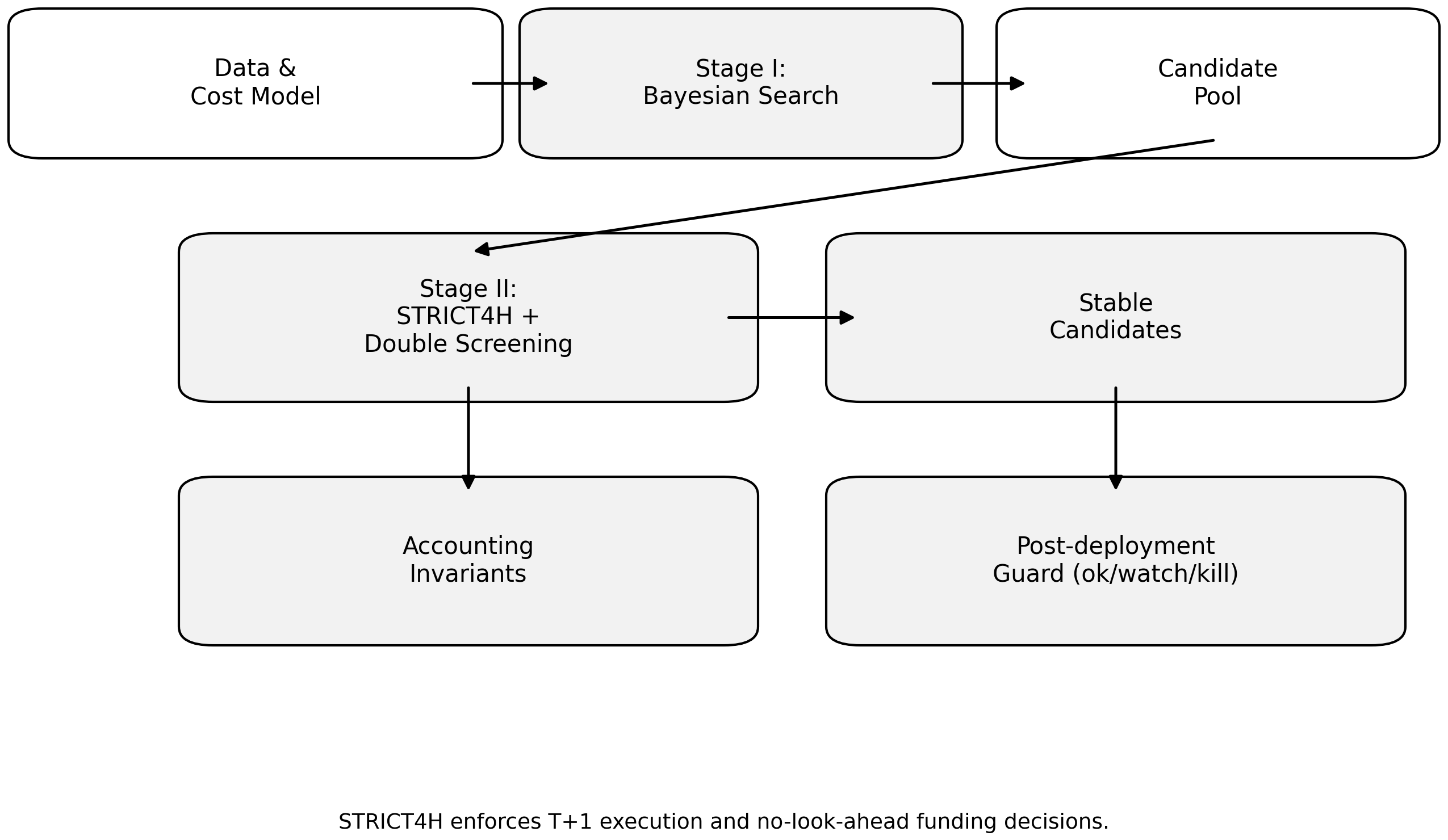}
  \caption{\textbf{AutoQuant methodology flowchart.} Stage I Bayesian search produces candidates that undergo Stage II double-screening; stable sets are supervised by a guard overlay. STRICT4H enforces \(t\!+\!1\) execution and no-look-ahead funding alignment.}
  \label{fig:flow_method}
\end{figure}

Figure~\ref{fig:flow_method} provides a high-level overview of the two-stage workflow and its traceability interfaces, including window splits, scenario grids, and stable-candidate selection.

\FloatBarrier

\subsection{Algorithmic Summary (Pseudo-code)}\label{sec:algo_summary}

To make the end-to-end procedure directly reproducible from the text, the pipeline is summarized as a minimal interface and pseudo-code.
Inputs are (i) a time-aligned 4-hour OHLCV series, (ii) an 8-hour funding-rate series aligned to the STRICT4H grid, (iii) a fixed cost profile and a predefined cost-scenario grid, (iv) a window split (Training/Validation/Long-horizon/Blind hold-out), and (v) a Stage~I search budget \(N_{\mathrm{opt}}\).
Outputs are the selected configuration, the backtest logs, and the exported machine-readable artifacts described in Appendix~A.

\begingroup
\footnotesize
\begin{verbatim}
# Stage I: Bayesian search (train net return)
history = []
for i in 1..N_opt:
    theta_i ~ TPE(history)
    r_net = STRICT4H_backtest(
        data=Training, theta=theta_i, costs=baseline_costs
    )
    score_i = annualized_return(r_net)
    history.append((theta_i, score_i))
Theta_pool = top-K candidates by score_i

# Stage II: double-screening (validation for selection only)
for theta in Theta_pool:
    for scenario in cost_grid:
        r_net = STRICT4H_backtest(
            data=Train+Val, theta=theta, costs=scenario
        )
        compute monthly return, drawdown, turnover
select theta_star via filters and scenario metrics

# Stress tests (blind hold-out; not used for selection)
evaluate theta_star on Long-horizon and Blind hold-out windows
export tabular summaries
\end{verbatim}
\endgroup

In the implementation, the ``export tabular summaries'' step is backed by deterministic machine-readable outputs, including configuration \texttt{json}, per-bar ledger \texttt{csv}, robustness summaries, and invariant-check files that support manuscript table/figure reconciliation and backtest-to-replay checks when the companion data artifacts are available, as summarized in Appendix~A.

\subsection{The Signal-Modular Parameter Space}\label{sec:theta_space}

The framework deliberately treats the underlying alpha-generation logic
as a black-box mapping
\[
  f : \mathcal{I}_t \times \Theta \rightarrow S_t,
\]
where \(\mathcal{I}_t\) denotes the information set available at time
\(t\), \(\Theta\) the hyperparameter space, and \(S_t \in \{-1,0,1\}\)
the target position or trading signal. In the STRICT4H case study, the
core momentum signal is derived from the normalized divergence between
fast and slow exponential moving averages of the 4-hour close,
compressed into \([0,1]\) via a smooth \(\tanh\) transform and combined
with auxiliary mean-reversion and breakout channels. The details of this
construction are implementation-specific; the focus here is on the structure
and constraints of \(\Theta\) rather than on the full functional form of
the alpha engine itself.

While the AutoQuant architecture is signal-agnostic, the specific
hyperparameter search space \(\Theta\) in this manuscript is optimized
for trend-following and mean-reversion families on cryptocurrency
perpetual futures. Extending the framework to high-frequency
market-making or arbitrage strategies would require redefining \(\Theta\)
and the associated cost-model primitives, although the validation
pipeline would remain conceptually unchanged. Importantly, AutoQuant does
not attempt to invent \(f\) automatically; the novelty lies in enforcing
strict semantics and providing an auditable selection-and-screening
pipeline for \(\theta \in \Theta\).

\begin{equation}
 m_{t}(\theta_{\mathrm{momentum}})
 = \frac{1}{2}\left[1 +
   \tanh\left(
     \frac{\mathrm{EMA}_{\mathrm{fast},t} - \mathrm{EMA}_{\mathrm{slow},t}}{
       \sigma(\theta_{\mathrm{momentum}})\,\mathrm{EMA}_{\mathrm{slow},t}}
   \right)
 \right].
\label{eq:momentum-score}
\end{equation}

\begin{equation}
 C_{t} = w_{\mathrm{mom}} \cdot m_{t} +
         (1 - w_{\mathrm{mom}}) \cdot a_{t}.
\label{eq:composite-score}
\end{equation}

\begin{equation}
\mathrm{Signal}_{t} =
\begin{cases}
  +1, & C_{t} \geq \tau_{t}^{\mathrm{long}},\\
  -1, & C_{t} \leq 1 - \tau_{t}^{\mathrm{short}},\\
  0,  & \text{otherwise}.
\end{cases}
\label{eq:signal-rule}
\end{equation}

Instead of enumerating implementation-specific variable names, the manuscript defines
the hyperparameter space \(\Theta\) in terms of a small number of
interpretable blocks. The momentum and mean-reversion block controls the
EMA lookback lengths, volatility-scaling factors, and Bollinger-band
widths and tail quantiles that determine how quickly the strategy reacts
to trends or reversals. The risk-control block specifies minimum holding
times, cooldown periods after exits, ATR-based stop-loss and take-profit
multipliers, and several exposure-smoothing and maximum-exposure
constraints. The Bayesian optimization stage explores \(\Theta\) within
fixed bounds chosen to respect realistic leverage, exposure, and cost
constraints.

\begin{table}[!b]
  \centering
  \small
  \setlength{\tabcolsep}{5pt}
  \renewcommand{\arraystretch}{1.12}
  \caption{\textbf{Minimal STRICT4H hyperparameter summary (\(\Theta\)) with default bounds and fallbacks.} Ranges correspond to the default Stage~I Optuna search space. ``Default'' denotes the fallback value used when a parameter is omitted.}
  \label{tab:theta_min}
  \begin{tabular}{@{}p{0.16\linewidth}p{0.24\linewidth}p{0.23\linewidth}p{0.31\linewidth}@{}}
    \toprule
    Block & Parameter & Range (default) & Constraint / interpretation \\
    \midrule
    Momentum & \texttt{ema\_fast} & 6--32 (12) & EMA span in 4h bars; must be shorter than \texttt{ema\_slow}. \\
    Momentum & \texttt{ema\_slow} & 20--96 (26) & EMA span in 4h bars; \texttt{ema\_fast} \(<\) \texttt{ema\_slow}. \\
    Momentum & \texttt{ema\_threshold} & 0.0005--0.0060 (0.003) & Entry threshold on EMA spread (decimal units). \\
    Mean-rev & \texttt{bb\_period} & 10--30 (20) & Bollinger lookback in 4h bars. \\
    Mean-rev & \texttt{bb\_dev} & 1.0--2.5 (2.0) & Band width in standard deviations. \\
    Risk & \texttt{min\_hold\_bars} & 1--6 (1) & Minimum holding duration in 4h bars. \\
    Risk & \texttt{cooldown\_hours} & 0--8 (0) & Cooldown after exits to reduce rapid flip-flopping. \\
    Risk & \texttt{atr\_period} & 0--30 (0) & ATR period in 4h bars; 0 disables ATR-based exits. \\
    Risk & \texttt{atr\_k\_sl}, \texttt{atr\_k\_tp} & 0--3 / 0--5 (0) & ATR stop-loss / take-profit multipliers; 0 disables. \\
    Risk & \texttt{max\_exposure\_abs} & 0--5 (0) & Account-level exposure cap. \\
    Funding & \texttt{funding\_bias\_\allowbreak thr\_bps} & 0--10 (0) & Soft threshold before raising entry thresholds under adverse funding. \\
    Funding & \texttt{funding\_bias\_\allowbreak k\_thr\_\allowbreak per\_bps} & 0--0.005 (0) & Threshold slope under adverse funding. \\
    \bottomrule
  \end{tabular}
\end{table}

To mitigate adverse funding costs, the framework introduces a soft funding-bias
mechanism that raises entry thresholds in periods of expensive carry. Let
\(fr_t\) denote the 8-hour funding rate aligned to the 4-hour grid via
carry-forward, and let \(\tau_{\mathrm{base}}\) be a baseline long-entry
threshold. For long positions, the effective threshold is defined as
\begin{equation}
  \tau_{\mathrm{long}}(t)
  = \tau_{\mathrm{base}} + \kappa \cdot
    \max\bigl(|fr_t| - \vartheta_{\mathrm{bias}}/10000, 0\bigr),
\label{eq:funding-threshold}
\end{equation}
where \(\vartheta_{\mathrm{bias}} \ge 0\) is an activation threshold and
\(\kappa \ge 0\) controls how quickly the threshold increases as funding
moves deeper into the tails.

The optimization process in Stage~I is guided by a single scalar
objective dominated by cost-adjusted annualized return. Formally,
\begin{equation}
  \max_{\theta \in \Theta} J(\theta), \qquad J(\theta) := r_{\mathrm{ann}}(\theta),
  \label{eq:stage1-objective}
\end{equation}
where \(r_{\mathrm{ann}}(\theta)\) is the annualized net return computed
from the STRICT4H net-return series under the Stage~I window and the
baseline cost profile.

\paragraph{Performance metrics}\label{sec:perf_metrics}
The primary performance metric, annualized return
\(r_{\mathrm{ann}}\), is based on geometric compounding of the 4-hour
strategy returns over the entire backtest period. Let \(R_{\mathrm{tot}}\)
denote the total compounded return over \(N\) 4-hour bars, and let \(AF\)
be the annualization factor (set to 2190):
\begin{equation}
 r_{\mathrm{ann}} = (1 + R_{\mathrm{tot}})^{AF/N} - 1.
\label{eq:annual-return}
\end{equation}
Unless otherwise stated, annualized return is reported as CAGR in decimal
units. Maximum drawdown (MaxDD) is reported as the peak-to-trough equity
decline over the window.

\subsection{Stage I: Bayesian Auto-Tuning under Realistic Costs}
The first stage of the framework employs Bayesian optimization to
efficiently search the parameter space. The implementation uses the Tree-structured Parzen
Estimator algorithm \citep{bergstra2011algorithms}, as implemented in
Optuna \citep{akiba2019optuna}, to guide the search.

\paragraph{Why TPE Relative to Grid Search, Evolutionary Methods, and GP-BO}
TPE is selected because the search space \(\Theta\) contains mixed
continuous/discrete parameters, conditional blocks, and hard feasibility
constraints. In this setting, TPE is a practical and widely used sampler
for conditional search spaces and performs well in compute-limited
regimes. By contrast, grid search scales poorly with dimension; random
search is a strong baseline under limited budgets; evolutionary methods
can be effective but often require larger evaluation budgets and more
careful constraint handling; and Gaussian-process Bayesian optimization
is less convenient for high-dimensional conditional spaces. Importantly,
AutoQuant is sampler-agnostic: any black-box optimizer, including genetic
algorithms, differential evolution, or GP-based Bayesian optimization, can be
substituted if it proposes \(\theta\) and consumes scalar scores. Such a
substitution would not change the strict execution/cost semantics or the
Stage~II screening policies. The present manuscript therefore justifies TPE as
a practical implementation choice rather than as an algorithmic contribution.
To assess sampler sensitivity, the empirical section also reports a same-budget
benchmark against random search, DE, GA, and GP-based Bayesian optimization.
That benchmark supports sampler substitutability under the same strict
evaluator; in the reported run, GP-BO finds the strongest training-window
candidate.

In the BTC/USDT STRICT4H case study reported in this manuscript, the Stage~I
search budget is fixed ex ante at \(N_{\mathrm{opt}}=40\) TPE trials on the
training window. Annualized net return is used as the Stage~I
scalar objective to keep the optimization problem simple and to avoid
embedding additional risk preferences into the sampler. Risk and
fragility controls are instead enforced in Stage~II as explicit,
auditable policy constraints. A key design choice is that Stage~I and
Stage~II evaluations share the same STRICT4H execution and
funding-alignment semantics; Stage~I differs only in its role, while
Stage~II re-evaluates a fixed candidate pool under additional windows and
cost-mis-specification scenarios.

\noindent\textbf{Optimizer baseline.}
As a minimal baseline under the same STRICT training window and objective,
the best-of-\(N\) curve of TPE is compared against random sampling
\citep{bergstra2012random} in Table~\ref{tab:optimizer_baselines}. This
baseline is used as a matched-budget diagnostic showing that the search component
is not merely an ad hoc selection device. Broader optimizer diagnostics are
reported in Section~\ref{sec:empirical}.
\begin{table}[!htbp]
  \centering
  \small
  \caption{\textbf{Optimizer baseline under identical STRICT training window.} Best-of-$N$ objective (annual). Random values are mean$\pm$std over seeds.}
  \label{tab:optimizer_baselines}
  \setlength{\tabcolsep}{6pt}
  \begin{tabular}{@{}rcc@{}}
    \toprule
    Budget $N$ & TPE best & Random best (mean$\pm$sd) \\
    \midrule
    10 & 0.0174 & -1.0000 $\pm$ 0.0000 \\
    20 & 0.0362 & -0.7832 $\pm$ 0.4848 \\
    40 & 0.1172 & -0.7832 $\pm$ 0.4848 \\
    60 & 0.1334 & -0.7832 $\pm$ 0.4848 \\
    80 & 0.1508 & -0.6638 $\pm$ 0.4915 \\
    100 & 0.1590 & -0.4600 $\pm$ 0.5163 \\
    120 & 0.1590 & -0.3580 $\pm$ 0.6763 \\
    \bottomrule
  \end{tabular}
\end{table}

\noindent\textbf{Compute budget disclosure.}
The dominant cost driver is the number of rigorous STRICT4H backtest
evaluations. Stage~I runs \(N_{\mathrm{opt}}\in\{40,120\}\) TPE trials on
the training window, with one STRICT4H backtest per trial.
Stage~II then re-evaluates a fixed candidate pool across a small
cost-scenario grid and multiple window segments. In the reproducible BTC
snapshot, the screening applies 9 cost scenarios to the top 40 candidates
over the combined Training+Validation window (7{,}262 4-hour bars),
totaling 360 strict backtest runs, plus a small number of long-horizon and
blind hold-out stress-test evaluations. All experiments are CPU-only. On a
representative workstation, end-to-end runtime is typically hours to under
one day per asset under these budgets.

\noindent\textbf{Supporting funding-rule diagnostic.}
To isolate whether funding-aware alignment rules matter beyond simply
charging funding as a cost, STRICT execution and all costs are kept fixed,
and only the funding-related gating and bias rules are disabled. Table
~\ref{tab:funding_gate_ablation} reports the resulting change in
performance for a representative tuned configuration. The diagnostic tests whether
funding awareness can change the decision path itself, not only the ex post
PnL calculation. Because this single component diagnostic is not a broad
module-ablation suite, the main empirical claims do not depend on it.
\begin{table}[htbp]
  \centering
  \footnotesize
  \caption{\textbf{Supporting funding-rule diagnostic (GP-BO candidate).} Costs and STRICT execution are held fixed; only funding-related gating, bias, and carry rules are disabled. The table reports component-level sensitivity.}
  \label{tab:funding_gate_ablation}
  \setlength{\tabcolsep}{3pt}
  \begin{tabular}{@{}lcccc@{}}
    \toprule
    Variant & Train monthly geom & Train MaxDD & Val monthly geom & Val MaxDD \\
    \midrule
    Full & 0.218 & 0.192 & 0.000 & 1.000 \\
    NoFundingGates & 0.219 & 0.192 & 0.000 & 1.000 \\
    \bottomrule
  \end{tabular}
\end{table}

\subsection{Stage II: Rigorous 4h Backtest and Double Screening}\label{sec:strict4h_execution}

The candidate parameter sets generated in Stage~I then enter a rigorous
double-screening stage. The goal is no longer to search for new
configurations, but to determine whether seemingly attractive candidates
remain credible once they are subjected to strict execution semantics,
explicit funding-visibility rules, and bounded stress scenarios. The Stage~II
engine is therefore designed both to remove common sources of backtest
inflation and to keep the offline evaluation path aligned with live-style
execution logic.

\noindent\textbf{Strict \(t\!+\!1\) execution.}
A signal generated at the close of 4-hour bar \(t\), using only information
available at that time, can affect position and profit-and-loss only from bar
\(t+1\) onward. This prohibits same-bar execution and closes an important
source of look-ahead bias.

\noindent\textbf{No-look-ahead funding model.}
Funding costs are applied bar by bar. Funding rates are treated as an
external time series aligned to the 4-hour grid via carry-forward, and the
backtest uses only the funding values visible at each bar timestamp, scaled
to bar duration. The engine is not allowed to backfill from future funding
updates.

To make the accounting explicit, let \(S_t \in \{-1,0,1\}\) denote the
directional signal computed at the close of bar \(t\) from \(\mathcal{I}_t\).
STRICT4H enforces a strict \(t\!+\!1\) delay by applying an executed signed
nominal exposure \(\pi_t\) during bar \(t\), derived from \(S_{t-1}\) after
ex-ante caps, smoothing, and other fixed risk controls. Given the
close-to-close market return \(r^{\mathrm{mkt}}_{t}\) in
Eq.~\eqref{eq:return-def}, the raw strategy return is
\(r^{\mathrm{raw}}_{t} = \pi_t \, r^{\mathrm{mkt}}_{t}\). Transaction costs are
applied on exposure changes at bar boundaries, \(\Delta \pi_t = \pi_t -
\pi_{t-1}\), with fees and slippage proportional to \(|\Delta \pi_t|\).
Funding costs accrue on notional exposure \(\pi_t\) using the aligned 8-hour
funding series scaled to 4-hour bars. The resulting per-bar net return is
\begin{equation}
  r^{\mathrm{net}}_t = r^{\mathrm{raw}}_t - C_{\mathrm{fee},t} - C_{\mathrm{slip},t} - C_{\mathrm{fund},t}.
  \label{eq:net-return-def}
\end{equation}
All reported performance metrics in the empirical section are computed from the
resulting \(\{r^{\mathrm{net}}_t\}\) series unless stated otherwise.

\begin{table}[t]
  \centering
  \small
  \setlength{\tabcolsep}{5pt}
  \caption{\textbf{Core STRICT4H symbols, units, and accounting bases.} Unless stated otherwise, returns and costs are in decimal units; bps inputs are converted by dividing by 10,000.}
  \label{tab:execution_symbols}
  \begin{tabular}{@{}p{0.15\linewidth}p{0.25\linewidth}p{0.54\linewidth}@{}}
    \toprule
    Symbol & Units / range & Meaning \\
    \midrule
    $S_t$ & $\{-1,0,1\}$ & Directional signal computed at the close of bar $t$. \\
    $\pi_t$ & unitless, bounded & Executed signed nominal exposure held during bar $t$, applied with a strict $t\!+\!1$ delay from $S_{t-1}$ after fixed caps, smoothing, and other risk controls. \\
    $r^{\mathrm{mkt}}_t$ & decimal & Close-to-close market return of the underlying over bar $t$. \\
    $C_{\mathrm{fee},t}, C_{\mathrm{slip},t}$ & decimal & Turnover-based trading costs applied at bar boundaries, proportional to $|\Delta \pi_t|$; bps inputs apply to notional turnover. \\
    $C_{\mathrm{fund},t}$ & decimal & Funding cost accrued on notional exposure $\pi_t$ using the aligned 8-hour funding-rate series scaled to 4-hour bars. \\
    \bottomrule
  \end{tabular}
\end{table}

Stage~II applies two validation streams to a fixed pool of Stage~I
candidates; there is no additional re-optimization in this phase.

\noindent\textbf{Held-out window robustness.}
Each candidate is re-evaluated across multiple overlapping rolling windows of
fixed length (2,000 bars in the BTC case study, stepped forward by 500 bars)
that are held out from Stage~I optimization but may be reused within
Stage~II. This differs from classical walk-forward optimization because the
parameter set is not re-trained on each window; instead, the same fixed
configuration is applied repeatedly to assess temporal stability.

\noindent\textbf{Cost-scenario robustness.}
The same parameter sets are re-evaluated under a bounded set of
cost-mis-specification profiles. For stable-candidate filtering, the framework uses a
discrete sensitivity grid that varies taker fees and funding multipliers
\((\mathrm{taker\_bps}\in\{3,4,6\}\) and
\(\mathrm{fund\_mult}\in\{0.5,1.0,1.5\})\), producing scenario-aggregated
monthly and drawdown summaries for each candidate. This grid includes both
optimistic and pessimistic perturbations around a conservative baseline to make
cost sensitivity explicit. As an additional stress test on selected
configurations, pessimistic-cost scaling scenarios are also reported that multiply
transaction-related costs by fixed factors (for example, \(2\times\) and
\(3\times\)) together with a funding-off stress.

Stable-candidate selection then proceeds in three transparent steps. First, absolute robustness thresholds are
applied to remove underperforming
configurations. In the BTC STRICT4H experiments these ex-ante rules include a
minimum average monthly return, a floor for the worst scenario-level monthly
return, a ceiling on average maximum drawdown, and a cap on
position-switch density. Second, the surviving candidates are ranked by
average monthly return in descending order and average maximum drawdown in
ascending order over the training-plus-validation regime. Third, the top \(K\) candidates are retained
as the final pool of stable BTC/USDT STRICT4H
configurations.

To make this policy transparent, Table~\ref{tab:stable_candidate_rules}
reports the exact thresholds used in the stable-candidate filter. In the
replication code these rules are implemented as an ex-ante filter over
scenario-aggregated performance summaries followed by a simple lexicographic
ranking by mean performance and drawdown; they are not tuned post hoc on a
per-experiment basis.

\begin{table}[htbp]
\centering
\footnotesize
\setlength{\tabcolsep}{4pt}
\caption{Stable-candidate selection rules used in the BTC/USDT STRICT4H case study. After filtering, the top \(K=5\) candidates are retained by sorting \(\mathrm{mean\_monthly\_true}\) descending and \(\mathrm{maxDD\_mean}\) ascending.}
\label{tab:stable_candidate_rules}
\begin{tabular}{@{}>{\raggedright\arraybackslash}p{0.25\linewidth}>{\raggedright\arraybackslash}p{0.42\linewidth}>{\raggedright\arraybackslash}p{0.13\linewidth}>{\raggedright\arraybackslash}p{0.14\linewidth}@{}}
\toprule
Statistic & Meaning & Threshold & Role \\
\midrule
\texttt{mean\_\allowbreak monthly\_\allowbreak true} & mean monthly geometric return (across scenarios) & \(\ge 0.005\) & minimum performance \\
\texttt{min\_\allowbreak monthly\_\allowbreak true} & worst-case scenario-level monthly geometric return & \(\ge 0.0\) & downside floor \\
\texttt{maxDD\_\allowbreak mean} & mean maximum drawdown (across scenarios) & \(\le 0.30\) & drawdown cap \\
\texttt{switch\_\allowbreak density\_\allowbreak mean} & mean position-switch density (across scenarios) & \(\le 0.12\) & turnover cap \\
\bottomrule
\end{tabular}
\end{table}

\noindent\textbf{Metric choice note.}
Stable candidates are ranked using monthly geometric mean, drawdown, and
switching frequency because these metrics are directly interpretable and less
distorted by extreme compounding in high-volatility regimes. Other stability
metrics, such as Sharpe-, Sortino-, or Calmar-style summaries, can be
substituted within the same policy layer, but a systematic comparison of
screening metrics is outside the scope of the present study.

\noindent\textbf{Stable-candidate statistics and aggregation.}
For a candidate parameter set and a cost scenario \(s\) from the predefined
grid, the net return series \(\{r^{\mathrm{net}}_t\}\) is computed and resampled
into calendar-month compounded returns
\(R_{m,s} = \prod_{t \in m}(1+r^{\mathrm{net}}_{t,s})-1\). The
scenario-level monthly geometric mean is
\[
g_s = \exp\!\left(\frac{1}{M}\sum_m \log(1+\max(R_{m,s},-0.999999))\right)-1 .
\]
This value is reported as \(\mathrm{monthly\_true}=g_s\). The stable-candidate
filter aggregates these scenario-level statistics across the scenario grid:
\[
\begin{aligned}
\mathrm{mean\_monthly\_true} &= \mathbb{E}_s[g_s],\\
\mathrm{min\_monthly\_true} &= \min_s g_s,\\
\mathrm{maxDD\_mean} &= \mathbb{E}_s[\mathrm{maxDD}_s].
\end{aligned}
\]
Position-switch density is computed per scenario as
\[
\mathrm{switch\_density}_s = \#\{\Delta \pi_t \ne 0\}/N,
\]
and the reported summary is
\(\mathrm{switch\_density\_mean}=\mathbb{E}_s[\mathrm{switch\_density}_s]\).
These definitions match the exported robust-summary files referenced in
Appendix~A.

\section{Related Literature}

This section positions AutoQuant relative to three literature streams that are
usually studied separately but interact tightly in frictional perpetual-futures
research: realistic backtesting under market frictions, hyperparameter search
and selection bias in quantitative trading, and deployment-oriented traceability
and governance for decision-support systems. This organization makes clear
that the contribution lies less in proposing a new predictive model than in
connecting these strands into an auditable configuration-selection protocol.

\subsection{Expert Systems, Decision Support, and Auditable Pipelines}
AutoQuant is positioned as an \emph{expert-system-like}, rule-based
decision-support pipeline: it codifies non-negotiable semantics and
constraints as a knowledge base, uses an inference procedure
(Bayesian search plus policy-driven screening) to select robust
configurations, and exports traceable artifacts
(Section~\ref{sec:algo_summary} and Table~\ref{tab:eswa_mapping}). This
emphasis is aligned with decision-support objectives in operational
fintech contexts, where governance depends on explicit measurement,
execution, and control layers rather than on opaque ``best backtest''
reports \citep{philippon2016fintech,gomber2017digitalfinance}. In this
sense, AutoQuant is closer to an execution-aware, traceability-oriented AutoML
workflow \citep{hutter2019automl} than to a strategy-generation system:
the signal family is fixed, and the contribution lies in making
configuration selection reproducible and inspectable under strict
\(t\!+\!1\) semantics and cost realism.

\subsection{Backtesting and Execution Biases in Cryptocurrency Markets}
Realistic backtesting is difficult even in traditional assets
\citep{lopez2018advances}, and the problem is amplified in cryptocurrency
perpetual futures. The 24/7 trading cycle and extreme volatility make data
alignment and timestamp integrity critical
\citep{liu2020riskscrypto,liu2022commonriskcrypto}. Prior work highlights
look-ahead bias, microstructure frictions, arbitrage constraints, and
liquidity fragmentation as recurring sources of inflated backtest evidence in
crypto markets \citep{makarov2020trading}. These findings motivate
strict execution assumptions rather than post-hoc adjustments.

More specific to perpetual contracts, funding modeling remains a major
source of error. Recent work on perpetual-futures pricing, basis risk, and
market quality shows that contract design and funding mechanics can be
first-order drivers of both profitability and liquidity, so backtests that
omit or misalign funding can be materially misleading
\citep{ackerer2024perpetualpricing,gornall2025fundingcrisisproofed,ruan2025perpetualquality}. A complementary systems-oriented review of
perpetual futures across centralized and decentralized exchanges likewise
shows how exchange design and trader behavior shape those mechanics
\citep{chen2024perpetualcexdex}. Recent evidence also documents strong
linkages between spot and derivatives markets, including price discovery in
Bitcoin futures \citep{hung2021pricediscovery,shynkevich2020infoeff}.
Empirical work on modeling and forecasting perpetual-futures prices for
trading applications further underscores that data construction and contract
settlement mechanics matter for inference
\citep{malik2023perpetualpricing}. This work builds on that literature by
programmatically enforcing \(t\!+\!1\) execution and no-look-ahead funding
calculations in the backtest engine, moving from problem identification to
a traceable implementation pattern.

\subsection{Parameter Tuning and Bayesian Optimization in Finance}
Most quantitative strategies are highly sensitive to parameter choices.
Manual tuning and naive grid search are inefficient and can promote
overfitting \citep{lopez2019seven}. Bayesian optimization and related
methods can navigate large search spaces more efficiently
\citep{snoek2012practical}.

Random search provides a competitive and widely used baseline for
hyperparameter optimization under limited evaluation budgets
\citep{bergstra2012random}, and evolutionary computation methods
(e.g., genetic algorithms, differential evolution, and CMA-ES) provide a
complementary family of black-box optimizers with different trade-offs in
constraint handling and sample efficiency
\citep{goldberg1989ga,storn1997de,hansen2006cmaes}.

In finance, Bayesian optimization has been applied to trading rules and
portfolio allocation, but a common limitation is that optimization is often
decoupled from realistic cost and risk modeling. Parameters that appear strong
in a simplified search environment can later prove fragile under real-world
frictions. This concern is closely related to classic data-snooping and
multiple-testing problems: repeated search over strategy variants or
hyperparameters can induce selection bias even when each backtest is
implemented correctly \citep{sullivan1999datasnooping,white2000realitycheck,hansen2005spa,harvey2015crosssection}. The use of DSR/PBO-style diagnostics
is therefore a robustness check and complements rather than replaces strict execution
and cost semantics \citep{bailey2014deflated,bailey2017pbo}. The contribution
here is to integrate Bayesian optimization with realistic costs from Stage~I
onward while keeping execution semantics fixed during Stage~II screening.

\subsection{Risk Management, Guards, and Full-Chain Consistency}
A robust quantitative trading system requires more than a profitable
signal; it also requires risk management and validation layers that remain
consistent from research to live deployment. Risk-based overlays, or
``guards,'' that monitor realized performance and can reduce or halt
trading are a familiar part of institutional practice. Recent work on
decentralized finance likewise treats protocol-level risk controls and
auditable accounting as first-class design objectives \citep{schar2021defi},
reinforcing the view that infrastructure and monitoring are integral to
decision support and model governance.

A parallel concern is the ``full-chain consistency'' between the backtest
environment and the live execution engine. Divergences in data handling,
cost calculation, or order logic can invalidate backtest results and lead
to unexpected live performance \citep{lopez2018advances}. This work
synthesizes these two concerns by proposing a framework that not only
includes a post-deployment guard but also enforces a set of
accounting invariants across the entire backtest-to-live chain. This
provides a formal, traceable link between theoretical research and
practical implementation, a critical but often overlooked aspect of
financial data science.

\section{Data and Market Setting}\label{data-and-market-setting}

This section outlines the financial instruments, data sources, and market assumptions that form the foundation of the empirical analysis. The study places a strong emphasis on data integrity and the realistic modeling of trading costs, since both are prerequisites for robust backtesting and transparent evaluation.

\subsection{Instruments and Data Sources}
The primary instrument for this study is the BTC/USDT perpetual futures contract, with parallel replications on ETH/USDT, SOL/USDT, and AVAX/USDT perpetual contracts using the same STRICT4H family and evaluation scripts. The experiments use 4-hour OHLCV bars sourced from major centralized-exchange perpetual-futures endpoints, with prices and volumes constructed from exchange-level candle fields and cleaned using the procedures described below. In the anonymized replication package, the BTC/USDT anchor series is implemented from Binance, Bybit, and OKX perpetual data, and the ETH/USDT, SOL/USDT, and AVAX/USDT replications are generated through the same venue-recorded download and cleaning scripts for the corresponding perpetual contracts. The cleaned series used in each experiment is a single timestamp-aligned perpetual-futures OHLCV series per asset rather than a volume-weighted cross-venue synthetic price. When more than one venue snapshot is available, the replication scripts select a single continuous per-asset venue stream after continuity and integrity checks rather than merging prices across venues; the exact local venue/source path is recorded in the regeneration configuration. The replication materials provide regeneration scripts and local processed-data paths, while raw exchange files are not redistributed. The dataset is partitioned into a Core Sample (2019-09-08--2021-12-31, \(N = 5{,}069\) 4h returns) used for descriptive statistics and core cost-accounting experiments, and an Extended Long-Horizon Sample (2019-09-08--2025-10-14, \(N = 13{,}368\) 4h returns) used for long-horizon diagnostics and robustness stress tests.

Crucially, the dataset includes a synchronized time series of funding rates extracted at 8-hour intervals, aligned to the 4-hour time axis as a step function carried forward between funding timestamps. The fallback is activated only when a realized funding timestamp is missing or unavailable for the relevant historical slice; in that case, the conservative constant funding-rate assumption from the same trading-cost profile is used and the carry-forward, no-backfill convention is kept unchanged.

For ETH/USDT, SOL/USDT, and AVAX/USDT, the study uses the same construction convention: one cleaned 4-hour OHLCV series and one corresponding 8-hour funding-rate series per asset, drawn from the venue-specific perpetual-futures records retained in the reproducible snapshot. Data availability differs by asset: the ETH series spans 2021-12-31--2025-11-24, while the SOL/AVAX series span 2021-01-01--2025-12-16. They are evaluated under the same strict execution semantics and a conservative cost family, using asset-specific fee/slippage baselines together with the Stage~II cost-sensitivity grid.

\begin{table*}[t]
  \centering
  \scriptsize
  \setlength{\tabcolsep}{3pt}
  \renewcommand{\arraystretch}{1.08}
  \caption{Data summary for BTC/USDT 4-hour returns.}
  \label{tab:btc4h_data}
  \begin{tabular}{@{}p{0.28\textwidth}rrrrr@{}}
    \toprule
    Sample & Returns (4h) & Mean & Std. dev. & Min & Max \\
    \midrule
    Core\\(2019-09-08--2021-12-31) & 5{,}069 & 0.00043 & 0.01615 & -0.21046 & 0.14734 \\
    Long\\(2019-09-08--2025-10-14) & 13{,}368 & 0.00026 & 0.01309 & -0.21046 & 0.14734 \\
    \bottomrule
  \end{tabular}
\end{table*}

In this manuscript, the label \emph{core} refers specifically to the BTC/USDT 4-hour segment used for the naive-versus-rigorous cost and constraint stress test. Separately, the BTC/USDT windowed experiments use explicit training/validation splits that extend through 2023-01-01, and long-horizon diagnostics extend through 2025-10-14.

Throughout the paper, date ranges are interpreted as half-open intervals
\([{\rm start},{\rm end})\), so adjacent windows do not share bars. The
following terminology is used for BTC/USDT 4-hour windows, with analogous
windows for the ETH/SOL/AVAX replications:
\begin{itemize}
  \item \emph{Training window} (2019-09-08--2021-01-01): used exclusively for Stage~I Bayesian optimization.
  \item \emph{Validation window} (2021-01-01--2023-01-01): used within Stage~II double-screening for robustness checks and selection, but not for Stage~I candidate generation.
  \item \emph{Long-horizon robustness window} (2019-09-08--2025-10-14): used for long-run diagnostics and cost-stress tests, not for parameter selection.
  \item \emph{Blind post-ETF window} (2024-01-01--2025-10-14): held out from tuning and selection and used solely for post-hoc robustness assessment.
\end{itemize}

All Stage~II thresholds and filters are fixed \emph{ex ante} prior to evaluation on the long-horizon or blind hold-out windows, and hyperparameters, thresholds, or selection rules are not modified after observing hold-out outcomes.

\begin{table*}[t]
  \centering
  \footnotesize
  \setlength{\tabcolsep}{3pt}
  \caption{Mapping between evaluation stages and BTC/USDT 4-hour windows in the AutoQuant pipeline.}
  \label{tab:stages_windows}
  \begin{tabular}{@{}>{\raggedright\arraybackslash}p{0.18\linewidth}>{\raggedright\arraybackslash}p{0.20\linewidth}>{\raggedright\arraybackslash}p{0.44\linewidth}>{\raggedright\arraybackslash}p{0.14\linewidth}@{}}
    \toprule
    Stage / component & Primary role & Windows used & Key settings \\
    \midrule
    Stage I (Bayesian search) & Parameter generation & Training & TPE; \(N_{\mathrm{opt}}=40\) \\
    Stage II (double-screening) & Robustness screening & Training + validation & \(K=5\); WFE win=2000, step=500 \\
    Long-horizon diagnostics & DSR and cost stress & Long-horizon robustness & cost stress \(2\times\), \(3\times\) \\
    Blind evaluation & Post-ETF robustness & Blind post-ETF & post-hoc only \\
    \bottomrule
  \end{tabular}
\end{table*}

Returns are computed as simple 4-hour arithmetic close-to-close returns,
\begin{equation}
 r^{\mathrm{mkt}}_{t} = \frac{P^{\mathrm{close}}_{t}}{P^{\mathrm{close}}_{t - 1}} - 1
\label{eq:return-def}
\end{equation}
where \(P^{\mathrm{close}}_t\) denotes the 4-hour close price. Under STRICT4H execution semantics, the position applied to \(r^{\mathrm{mkt}}_t\) is lagged by one bar, so that the raw strategy return at bar \(t\) depends only on information available by the close of bar \(t-1\).

\subsection{Data Cleaning and Validation}\label{sec:data_validation}
The policy for handling OHLCV data gaps is conservative and designed to prevent the introduction of artificial price data. The system does not interpolate or fill missing OHLCV bars. Funding-rate timestamps are treated as a separate external series: when available, they are aligned to the 4-hour grid using carry-forward between funding updates with no backfilling from future updates; when unavailable, the system falls back to a conservative constant assumption from the cost profile under the same carry-forward semantics.

\subsection{Data Construction Protocol (4-hour OHLCV)}
To make the BTC/USDT 4-hour series traceable, the study adopts a simple and explicit construction protocol: use exchange-provided perpetual-futures candle fields as the primary price/volume source; represent each bar by its bar-open timestamp in UTC; enforce basic OHLCV sanity constraints; sort and de-duplicate by timestamp; run a strict continuity check at the target frequency; and construct the STRICT4H funding series separately as an external input with an explicit fallback when realized funding is unavailable.

In the unified codebase used for this manuscript, the data-fetch and merge logic is implemented in lightweight scripts that support direct inspection. The replication package includes scripts for fetching full-history K-lines, aggregating 1-hour bars into 4-hour OHLCV, validating continuity and data integrity, and generating or backfilling the aligned funding series described above.

\section{System Architecture and Implementation}\label{sec:architecture}

This section summarizes the implementation choices that are most relevant
to the traceability of the empirical results under strict execution and
cost modeling. The full AutoQuant codebase contains additional production
components, including deployment utilities, exchange adapters, and
monitoring dashboards. This manuscript focuses on the design choices
that bear directly on research validity: (i) reuse of the same STRICT4H
signal and cost engine across offline evaluation and live-style replay;
(ii) explicit, reconcilable accounting of returns and costs; and (iii) a
post-deployment guard that is intentionally separated from the
optimization objective.

\subsection{Three Evaluation Stages: Data, Optimization, and Out-of-Sample Validation}
The workflow is organized into three stages. First, \emph{data and
environment validation} performs the integrity checks in
Section~\ref{sec:data_validation} and fixes a cost profile and constraint
set that remain unchanged during optimization and screening. Second,
\emph{optimization and screening} implements the two-stage methodology:
Stage~I generates candidates on a single training window under the
realistic cost model, while Stage~II screens candidates across held-out
windows and cost-stress scenarios using the STRICT4H engine. Third,
\emph{live-style replay and risk control} applies the same STRICT4H logic
on a rolling window of recent data to produce a target exposure and
optionally overlays a guard for post-deployment supervision. In this
paper, Stage~(iii) is treated as a traceability pattern rather than as
evidence of long-horizon live performance.

\subsection{The Live Engine and Executor}
The live engine is implemented as a thin wrapper around the same core
STRICT4H strategy and cost logic used offline. On each update cycle it
recomputes the target exposure from a bounded sliding window of recent
4-hour data and applies the same strict \(t\!+\!1\) semantics described in
Section~\ref{sec:strict4h_execution}. The executor then maps that target
exposure to exchange orders while enforcing the same leverage and notional
caps used in the backtest engine and logging fills and cash-flow
components, including fees, slippage, funding, and PnL, for later
reconciliation.

\subsection{Full-Chain Invariants and Guard Infrastructure}
A set of full-chain accounting invariants is defined: mathematical
identities that must hold, within a small tolerance, across the entire
lifecycle of a strategy instance. For any given parameter set, key
metrics such as cumulative exposure, total fees, total slippage, net
funding costs, and total PnL must reconcile between the offline
backtest, live-style replay, and execution logs produced by the same code
path when such logs are available.

Concretely, for each bar \(t\), the system records the raw strategy return
\(r^{\mathrm{raw}}_t\) and cost components \(C_{\mathrm{fee},t}\),
\(C_{\mathrm{slip},t}\), and \(C_{\mathrm{fund},t}\), and defines the net
return
\begin{equation}
  r^{\mathrm{net}}_t = r^{\mathrm{raw}}_t - C_{\mathrm{fee},t} - C_{\mathrm{slip},t} - C_{\mathrm{fund},t}.
\end{equation}
Full-chain consistency checks verify that these components reconcile across offline
backtests, live-style replay, and execution logs produced by the same
code path.

The repository-level deployment-bridge evidence is a strict-\(t\!+\!1\)
code-path parity check between the rigorous
backtest engine and the live-style inference wrapper. Across three independent
SOL/USDT 1-hour parameter snapshots over 2{,}000 bars each, the exported
signal and exposure series matched exactly. This is reported as internal consistency
evidence for the offline-to-live bridge rather than as evidence of
long-horizon live performance.

\begin{table}[htbp]
  \centering
  \small
  \setlength{\tabcolsep}{4pt}
  \caption{\textbf{Repository-level strict-\(t\!+\!1\) parity snapshots (rigorous backtest vs replay path).} Values are intended as code-path consistency evidence, not as a live-performance claim.}
  \label{tab:guard_audit}
  \begin{tabular}{@{}llrrrr@{}}
    \toprule
    Snapshot & Asset / freq & Bars & Direction match & $\max|\Delta S|$ & $\max|\Delta \pi|$ \\
    \midrule
    Trial 37 & SOL/USDT 1h & 2000 & 1.0 & 0 & 0 \\
    Trial 71 & SOL/USDT 1h & 2000 & 1.0 & 0 & 0 \\
    Trial 82 & SOL/USDT 1h & 2000 & 1.0 & 0 & 0 \\
    \bottomrule
  \end{tabular}
\end{table}

A guard provides a post-deployment risk overlay that monitors
rolling performance and risk metrics computed from execution logs and can
issue an \emph{ok/watch/kill} decision. A ``kill'' decision triggers
position flattening and can optionally disable further trading until
manual intervention. Because the guard responds to realized PnL and risk
states rather than to predictive signals, it is treated as a safety layer
that is conceptually separate from the Stage~I/II optimization
objectives. This paper reports only limited offline and replay-style
consistency evidence and therefore interprets the guard component as a
traceability pattern rather than as a live-performance claim.

\subsection{Computational complexity and deployment feasibility}
\label{sec:complexity_feasibility}

To make the practical footprint of AutoQuant explicit, the analysis distinguishes the
computational burden of Stage~I search, Stage~II screening, and online
deployment-style replay. Let \(T\) denote the number of bars in the evaluation
window, \(N_{\mathrm{opt}}\) the Stage~I search budget, \(K\) the number of
candidates retained for Stage~II, and \(S\) the number of cost-stress
scenarios. Under a fixed signal family, one strict backtest pass is linear in
\(T\) up to small constant factors induced by the enabled signal channels and
risk controls. Stage~I therefore scales approximately as
\(\mathcal{O}(N_{\mathrm{opt}} T)\), while the dominant Stage~II screening cost
scales approximately as \(\mathcal{O}(K S T)\) over the selected windows. In
the reproducible case study these workloads remain practical on CPU-only
hardware because the study deliberately uses bounded trial budgets, a small
stress grid, and a fixed candidate pool rather than an open-ended nested search.

The implementation is also designed with deployment feasibility in mind. The
same rigorous engine used for offline strict backtests is reused for live-style
inference, so the offline and online paths do not rely on separate strategy
logic. In the online wrapper, only a bounded recent history is retained,
repeated polling on an unchanged last bar is short-circuited by a lightweight
tail-bar fingerprint, and funding inputs are aligned under the same visibility
rules used offline. This design does not constitute a claim of
institutional-scale latency-sensitive execution, but it does show that the
proposed framework is operationally feasible as a research-to-deployment
decision-support pipeline for medium-frequency perpetual-futures strategies
under explicit execution semantics.

\section{Empirical Evaluation}\label{sec:empirical}

This section evaluates AutoQuant as an execution-aware decision-support
pipeline. The empirical analysis first compares optimizer behavior and
component ablations under the same STRICT4H evaluator, then examines funding
and screening sensitivity, and finally tests whether the validation protocol
remains informative across additional assets, an auxiliary signal family, and
third-party replay paths. Together, these experiments show how explicit
execution rules, cost realism, and auditable screening change the
interpretation of backtest results in frictional perpetual-futures markets.

\paragraph{Empirical roadmap}
The empirical section is organized to answer six distinct questions: how
several mainstream optimizers behave under the same strict-evaluation budget,
which decision and screening modules materially affect selected candidates,
whether funding-aware decision rules matter beyond ex post funding charges,
whether selected configurations remain interpretable under inferential and
policy-sensitivity checks, whether the validation protocol is portable across
several liquid perpetual contracts under identical strict semantics, and
whether a bounded auxiliary out-of-family strategy family survives the same screening
rules.

\subsection{Experimental Setup}
The experiments use 4-hour OHLCV datasets for BTC/USDT, ETH/USDT,
SOL/USDT, and AVAX/USDT perpetual contracts, built from the reproducible
dataset described in Section~\ref{data-and-market-setting}. For the primary
comparative experiment in Section~\ref{sec:one_vs_two}, the core BTC sample is
supplemented with the extended 4-hour series, and three distinct,
non-overlapping windows are defined: an in-sample training window used exclusively for
Stage~I Bayesian optimization, a validation-and-selection window used within
the Stage~II double-screening protocol, and a blind test window used only for
hold-out robustness assessment. For BTC, the
2023-01-01--2024-01-01 interval is intentionally left unused so that the validation-era selection
window remains clearly separated from the post-ETF hold-out.

For ETH/USDT, the study design is mirrored with a training window
(2021-12-31--2023-01-01), validation window (2023-01-01--2024-01-01),
and post-2024 hold-out (2024-01-01--2025-11-24), using the same STRICT4H
configuration family and evaluation scripts. For SOL/USDT and AVAX/USDT,
the same training and validation definitions as ETH are used, and the
post-2024 hold-out is extended through 2025-12-16.

\begin{table}[!b]
  \centering
  \small
  \setlength{\tabcolsep}{4pt}
  \caption{\textbf{Configuration naming used in the manuscript.} The text distinguishes \emph{baseline} (a frozen reference from Stage~I), \emph{one-stage} (top-1 on training), and \emph{selected two-stage} (stable-candidate screening). ``Naive'' and ``standard'' are diagnostic stress-test variants applied to a separate stress-test configuration.}
  \label{tab:config_naming}
  \begin{tabular}{@{}p{0.18\linewidth}p{0.34\linewidth}p{0.18\linewidth}p{0.24\linewidth}@{}}
    \toprule
    Name in text & Definition & Used for selection? & Where referenced \\
    \midrule
    Baseline STRICT4H & Frozen reference built from top-ranked Stage~I parameters; evaluated across windows & No & Table~\ref{tab:strict4h_windows}, Fig.~\ref{fig:baseline_windows} \\
    One-stage (top-1) & Best Stage~I trial by training-window objective & Yes (Stage~I only) & Section~\ref{sec:one_vs_two} \\
    Selected two-stage & Stable candidate selected by ex-ante filter and scenario-grid aggregation & Yes (Stage~II) & Section~\ref{sec:one_vs_two} \\
    Standard (fee-only) & Diagnostic backtest with taker fees only; same strict execution & No & Section~\ref{sec:naive_vs_rigorous} \\
    Naive (zero-cost) & Upper bound with fees/slippage/funding removed (and optional cap relaxation) & No & Section~\ref{sec:naive_vs_rigorous} \\
    \bottomrule
  \end{tabular}
\end{table}
\FloatBarrier

To provide a point of comparison, the manuscript defines a Baseline Strategy
configuration. This is not a toy example, but a realistic
production-oriented baseline built from the top-ranked Stage~I parameter
sets for the BTC/USDT STRICT4H family and then frozen for all subsequent
window and stress-test evaluations. It serves as a fixed reference
configuration for inspection and ablation.

\begin{figure}[htbp]
  \centering
  \includegraphics[width=\linewidth]{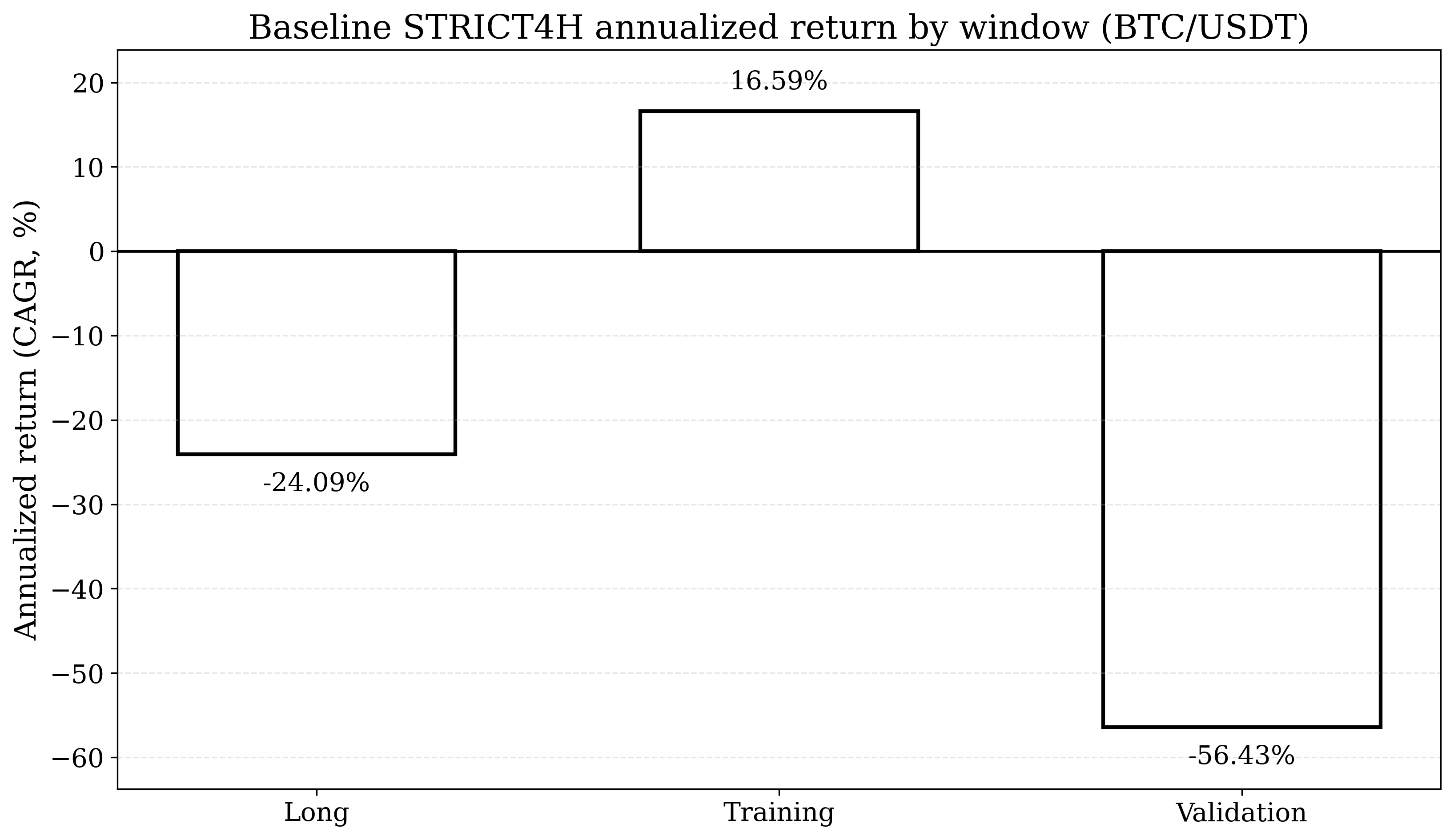}
  \caption{Baseline STRICT4H annualized return by window (BTC/USDT).}
  \label{fig:baseline_windows}
\end{figure}

\begin{table}[!htbp]
  \centering
  \small
  \setlength{\tabcolsep}{5pt}
  \caption{Baseline STRICT4H performance by window (BTC/USDT; annualized return reported as CAGR, \%; window definitions follow Section~\ref{data-and-market-setting}). Values are reported from the companion replication summaries described in Appendix~A.}
  \label{tab:strict4h_windows}
  \begin{tabular}{@{}>{\raggedright\arraybackslash}p{0.44\columnwidth}rrrr@{}}
    \toprule
    Window & Ann.\ (CAGR, \%) & MaxDD & Sharpe & Trades \\
    \midrule
    Long (2019-09-08--2025-10-14) & -24.09\% & 0.9536 & -0.0738 & 4{,}814 \\
    Training (2019-09-08--2021-01-01) & 16.59\% & 0.6041 & 0.5233 & 1{,}027 \\
    Validation (2021-01-01--2023-01-01) & -56.43\% & 0.9174 & -0.5913 & 1{,}549 \\
    \bottomrule
  \end{tabular}
\end{table}

In keeping with the emphasis on interpretable and economically meaningful
performance metrics, Table~\ref{tab:strict4h_windows} reports only
normalized statistics: annualized return, maximum drawdown, Sharpe ratio,
and total trade count, reported from STRICT4H machine-readable summaries in
the replication package. The same outputs also contain raw total
return and final-equity metrics, which are treated as internal accounting checks
and are deliberately omitted from the main text because they are highly sensitive to
sample length and the arbitrary choice of initial capital.

The Sharpe ratio is annualized from the 4-hour returns using the
annualization implied by the 4-hour frequency and assumes a constant annual
risk-free rate of 3\%. Because 4-hour crypto returns are autocorrelated and
heavy-tailed, Sharpe is treated as a descriptive statistic rather than as a
parametric test. The analysis therefore focuses not only on absolute metric
levels, but also on their sensitivity across time periods and cost
assumptions.

\noindent\textbf{Reading guide for compounded return magnitudes.}
Depending on the experiment, annualized return is reported either as CAGR in
\% or as CAGR in decimal units (e.g., 0.10 = 10\%). In high-volatility
regimes, compounded CAGR values can exceed 1.0 even under strict execution
and explicit leverage caps. Large CAGRs are therefore treated as
\emph{diagnostic} signals of sensitivity to execution, window choice, and
costs, and the interpretation places greater weight on drawdown, stability across windows,
and cost-stress outcomes. None of the reported return magnitudes should be
interpreted as statements about institutional-scale deployability, since
non-linear market impact is not modeled.

\subsection{One-Stage Tuning vs.~Two-Stage Double Screening}\label{sec:one_vs_two}
The first experiment is designed to test the core hypothesis of the
paper: that a simple one-stage parameter optimization is insufficient for
identifying robust strategies. Two distinct approaches are compared.

\noindent\textbf{One-Stage Tuning:} Stage~I Bayesian optimization is run
exclusively on the training window to find the top-performing parameter
set based solely on its in-sample, cost-adjusted annualized return.

\noindent\textbf{Two-Stage Framework:} The pool of
high-performing candidates from Stage~I is subjected to the full
Stage~II double-screening protocol, including out-of-sample rolling-window
analysis and cost-mis-specification scenarios. The final selected
parameter set is the one that ranks highest according to the
stable-candidate criteria described in Section~\ref{sec:strict4h_execution}.

Table~\ref{tab:one_vs_two_stage_btc} makes the BTC/USDT anchor comparison
explicit. Monthly geometric means are reported rather than headline CAGR in the
main table because the latter can become visually dominant in
high-volatility, small-sample regimes. The comparison is diagnostic: one-stage
tuning can deliver stronger in-sample and validation return magnitudes, while
the two-stage protocol makes the return--drawdown trade-off, turnover
differences, and hold-out degradation visible under unchanged strict
semantics. Buy-and-hold is also included as a simple external yardstick on the
same OHLCV series.

\begin{table*}[htbp]
  \centering
  \footnotesize
  \setlength{\tabcolsep}{5pt}
  \caption{\textbf{Direct comparison: one-stage versus two-stage selection under identical STRICT4H execution and rigorous costs (BTC/USDT).} One-stage selects the single best Stage~I trial on the training window. Two-stage selects a stable candidate via the ex-ante filter and cost-scenario aggregation described in Section~\ref{sec:strict4h_execution}. Validation is held out from Stage~I but reused within Stage~II screening; Post-2024 is a blind hold-out used only for robustness. The table reports monthly geometric net returns (decimal units), Sharpe, drawdown, and trade counts. Buy-and-hold is included as a simple external yardstick on the same OHLCV series.}
  \label{tab:one_vs_two_stage_btc}
  \begin{tabular}{@{}llrrrr@{}}
    \toprule
    Method & Window & Monthly geom & Sharpe & MaxDD & Trades \\
    \midrule
    One-stage & Training & 0.222 & 3.164 & 0.283 & 548 \\
    Two-stage & Training & 0.145 & 2.388 & 0.265 & 485 \\
    Buy-hold & Training & 0.067 & 1.530 & 0.563 & \textemdash \\
    One-stage & Validation & 0.051 & 1.238 & 0.318 & 423 \\
    Two-stage & Validation & 0.038 & 1.070 & 0.231 & 348 \\
    Buy-hold & Validation & -0.024 & 0.005 & 0.771 & \textemdash \\
    One-stage & Post-2024 & 0.023 & 0.845 & 0.220 & 259 \\
    Two-stage & Post-2024 & 0.016 & 0.664 & 0.196 & 223 \\
    Buy-hold & Post-2024 & 0.048 & 1.405 & 0.307 & \textemdash \\
    \bottomrule
  \end{tabular}
\end{table*}

\paragraph{Interpretation relative to the BTC anchor case}
The BTC anchor comparison highlights the selection trade-off directly. The
two-stage protocol trails the one-stage winner on return magnitude and
buy-and-hold in the post-2024 window, but under the same strict execution and
cost semantics it selects a lower-turnover candidate with shallower drawdowns
in validation and hold-out. This is the intended screening function: the protocol
makes fragility and selection trade-offs visible before deployment.

\noindent\textbf{Block-bootstrap check (validation window).}
As a minimal inferential supplement that is robust to time dependence, the
analysis computes a moving block bootstrap \citep{kunsch1989bootstrap} on the monthly
return series for the validation segment, reporting a 95\% interval for
the difference in monthly geometric mean returns between the two-stage and
one-stage procedures (two-stage minus one-stage).

\begin{table}[htbp]
  \centering
  \small
  \setlength{\tabcolsep}{5pt}
  \caption{\textbf{Moving block bootstrap CI for monthly geometric mean differences (BTC/USDT validation).} Monthly geometric means are in decimal units; \(\Delta\) is reported as two-stage minus one-stage.}
  \label{tab:bootstrap_one_vs_two_btc}
  \begin{tabular}{@{}lrrr@{}}
    \toprule
    Window & \(\Delta\) monthly geom (dec.) & 95\% CI (dec.) & Block length (months) \\
    \midrule
    Validation & -0.016 & [-0.040, 0.008] & 3 \\
    \bottomrule
  \end{tabular}
\end{table}

\paragraph{Threshold sensitivity}
Because stable-candidate selection relies on ex-ante thresholds, the replication package reports
a sensitivity scan in the replication package. Across reasonable
variations of the monthly return floor, drawdown cap, and switch-density
cap, the selected trial identity and its validation performance can
change, highlighting that these thresholds function as
\emph{risk-control design choices} rather than as universal constants. The
stable-candidate filter is therefore treated as a transparent, auditable
policy rather than as an optimized meta-learner. In the \(3\times 3\times
3\) grid (27 policy variants), the selected configuration toggles among
three trial IDs (52, 87, 108; each selected 9 times), with the number of
passing candidates ranging from 12 to 31. Across these policy variants,
validation-window CAGR spans 0.332--1.058 (decimal), Sharpe spans
0.844--1.667, and MaxDD spans 0.209--0.386. Table
~\ref{tab:threshold_sensitivity} shows representative settings.

\begin{table}[htbp]
  \centering
  \scriptsize
  \setlength{\tabcolsep}{3pt}
  \renewcommand{\arraystretch}{1.10}
  \caption{\textbf{Representative threshold-sensitivity settings for stable-candidate selection (BTC/USDT).} Thresholds are ex-ante design choices; the table reports the resulting selected trial ID and its validation performance under rigorous costs. Annual return is CAGR in decimal units; MaxDD is a fraction in \([0,1]\).}
  \label{tab:threshold_sensitivity}
  \begin{tabular}{@{}rrrrrrrr@{}}
    \toprule
    \shortstack{Monthly\\floor} & \shortstack{MaxDD\\cap} & \shortstack{Switch\\cap} & Selected & \shortstack{Val\\CAGR\\(dec.)} & \shortstack{Val\\Sharpe} & \shortstack{Val\\MaxDD} & \shortstack{Val\\trades} \\
    \midrule
    0.003 & 0.25 & 0.10 & 87  & 0.332 & 0.844 & 0.386 & 438 \\
    0.003 & 0.30 & 0.10 & 108 & 0.572 & 1.070 & 0.231 & 348 \\
    0.003 & 0.35 & 0.10 & 52  & 1.058 & 1.667 & 0.209 & 467 \\
    \bottomrule
  \end{tabular}
\end{table}

As an additional stress-test exercise, selected configurations are kept
fixed and extended into the 2021-12-31--2025-10-14 BTC/USDT 4-hour
period, including a blind post-ETF hold-out window
(2024-01-01--2025-10-14). These extensions are treated as post-hoc
robustness checks rather than as headline results, and they are not used
for parameter selection or for the core claims of the paper.

\subsection{Naive Backtest vs.~Rigorous Backtest}\label{sec:naive_vs_rigorous}
The second experiment quantifies the degree of performance overestimation
that results from extreme simplifications to the cost and constraint
model. For this purpose, the study constructs a deliberately optimistic
``Naive'' configuration that holds the STRICT4H signal logic and
\(t\!+\!1\) execution semantics fixed but removes market frictions and
relaxes capacity constraints. Relative to the rigorous configuration, the
naive configuration (i) sets fees, slippage, and funding costs to zero;
and (ii) relaxes leverage and notional caps so that the same signal path
can be evaluated in a near-frictionless environment. This should be
interpreted as an upper bound on friction- and constraint-related
performance inflation in this setting, rather than as a representation of
typical industry backtesting practice.

The analysis next isolates the impact of cost modeling itself. For this purpose, it
takes a \textbf{separate STRICT4H configuration} used specifically for this
stress test and evaluates it under a ladder of increasingly optimistic
assumptions on the BTC/USDT core sample. The \emph{rigorous} variant uses
the full realistic trading-cost profile under strict \(t\!+\!1\)
execution. The \emph{standard} variant keeps strict execution but
includes fees only. The \emph{naive} variant removes fees, slippage, and
funding to form a zero-cost upper bound. Because cost-aware channels
depend on the cost and funding inputs, the realized trade sequence and
thus the trade count can differ across variants even when the
hyperparameters are fixed.

\begin{table}[htbp]
  \centering
  \small
  \setlength{\tabcolsep}{4pt}
  \caption{\textbf{Cost ablation ladder for the naive-versus-rigorous stress test (BTC/USDT core sample).} Fixed parameters; the realized trade sequence may differ across variants. Annual return is reported as CAGR in decimal units. Values may exceed 1.0 under compounding and should be interpreted as a diagnostic of cost sensitivity rather than as a scalability claim. Market impact is not modeled.}
  \label{tab:cost_ablations}
  \begin{tabular}{@{}lrrrr@{}}
    \toprule
    Variant & Ann.\ (CAGR, dec.) & Sharpe & MaxDD & Trades \\
    \midrule
    Rigorous (fees+slippage+funding) & 2.726 & 2.049 & 0.265 & 636 \\
    Standard (fee-only) & 4.308 & 2.506 & 0.262 & 636 \\
    Naive (zero-cost, upper bound) & 5.225 & 2.711 & 0.260 & 535 \\
    \bottomrule
  \end{tabular}
\end{table}

Table~\ref{tab:cost_ablations} quantifies how optimistic cost
simplifications inflate reported performance for this fixed stress-test
configuration. Even a fee-only ``standard'' backtest can overstate
performance relative to a fully costed run that includes slippage and
funding, and removing all costs provides an additional optimistic upper
bound. The table is reconciled to the companion cost-accounting
ablation table in the replication materials, but it is used
only as a diagnostic cost-sensitivity ladder. These results are treated as
diagnostic baselines rather than as evidence of scalable profitability.

\subsection{Robustness and Cross-Asset Replication}
The final set of experiments evaluates the robustness of the parameter
sets identified by the full two-stage framework under alternative
conditions and across additional assets. First, the one-stage versus two-stage comparison is repeated
on ETH/USDT using the same STRICT4H family and identical execution and
cost semantics. Similar to the BTC anchor case, the two-stage selection makes
the trade-offs explicit under unchanged semantics: the selected candidate uses
far fewer trades and smaller drawdowns in two of the three windows, while the
one-stage top-1 retains higher return magnitude. This supports pipeline-level
portability of the screening mechanism across two liquid
perpetual contracts.

\begin{table}[htbp]
  \centering
  \small
  \setlength{\tabcolsep}{4pt}
  \caption{\textbf{Direct comparison: one-stage versus two-stage selection under identical STRICT4H execution and rigorous costs (ETH/USDT).} The table reports monthly geometric net returns (decimal units), Sharpe, drawdown, and trade counts.}
  \label{tab:one_vs_two_stage_eth}
  \begin{tabular}{@{}llrrrr@{}}
    \toprule
    Method & Window & Monthly geom (dec.) & Sharpe & MaxDD & Trades \\
    \midrule
    One-stage (top-1) & Training & 0.233 & 3.009 & 0.317 & 2112 \\
    Two-stage (selected) & Training & 0.173 & 2.607 & 0.276 & 346 \\
    One-stage (top-1) & Validation & 0.085 & 2.157 & 0.186 & 2012 \\
    Two-stage (selected) & Validation & 0.046 & 1.368 & 0.201 & 457 \\
    One-stage (top-1) & Post-2024 & 0.151 & 2.590 & 0.381 & 3880 \\
    Two-stage (selected) & Post-2024 & 0.096 & 2.009 & 0.368 & 852 \\
    \bottomrule
  \end{tabular}
\end{table}

Second, to probe higher-friction mid-cap contracts, the same protocol is
repeated on SOL/USDT and AVAX/USDT using stricter cost profiles but
the same strict execution semantics. The results illustrate the
intended role of double-screening as a \emph{fragility filter}: the
one-stage top-1 can exhibit extreme compounding and large drawdowns,
whereas the two-stage selection more often shifts the choice toward
lower-drawdown or less extreme configurations under stricter cost
assumptions. The SOL/USDT case is a useful counterpoint because the
two-stage candidate also improves several return metrics; this is treated as
a market- and candidate-specific outcome rather than a general rule. The
extreme AVAX/USDT one-stage compounding is interpreted as a volatility and
fragility diagnostic under linear costs.

\begin{table*}[t]
  \centering
  \footnotesize
  \setlength{\tabcolsep}{3pt}
  \caption{\textbf{Direct comparison on mid-cap contracts under stricter cost profiles (SOL/USDT and AVAX/USDT).} The table reports monthly geometric net returns (decimal units), Sharpe, drawdown, and trade counts.}
  \label{tab:one_vs_two_stage_midcaps}
  \resizebox{\textwidth}{!}{%
  \begin{tabular}{@{}llrrrrr@{}}
    \toprule
    Asset & Method & Window & Monthly geom (dec.) & Sharpe & MaxDD & Trades \\
    \midrule
    SOL/USDT & One-stage (top-1) & Training & 0.141 & 2.625 & 0.319 & 1567 \\
    SOL/USDT & Two-stage (selected) & Training & 0.124 & 3.013 & 0.152 & 1605 \\
    SOL/USDT & One-stage (top-1) & Validation & 0.227 & 3.398 & 0.239 & 1596 \\
    SOL/USDT & Two-stage (selected) & Validation & 0.362 & 4.208 & 0.242 & 1653 \\
    SOL/USDT & One-stage (top-1) & Post-2024 & 0.066 & 1.747 & 0.243 & 3295 \\
    SOL/USDT & Two-stage (selected) & Post-2024 & 0.068 & 2.301 & 0.195 & 3450 \\
    \midrule
    AVAX/USDT & One-stage (top-1) & Training & 0.479 & 3.239 & 0.795 & 2063 \\
    AVAX/USDT & Two-stage (selected) & Training & 0.091 & 4.856 & 0.055 & 1739 \\
    AVAX/USDT & One-stage (top-1) & Validation & 0.572 & 3.851 & 0.621 & 1908 \\
    AVAX/USDT & Two-stage (selected) & Validation & 0.087 & 4.185 & 0.096 & 1118 \\
    AVAX/USDT & One-stage (top-1) & Post-2024 & 0.797 & 4.407 & 0.607 & 4158 \\
    AVAX/USDT & Two-stage (selected) & Post-2024 & 0.094 & 3.994 & 0.142 & 3663 \\
    \bottomrule
  \end{tabular}
  }
\end{table*}

\begin{figure}[htbp]
  \centering
  \includegraphics[width=\linewidth]{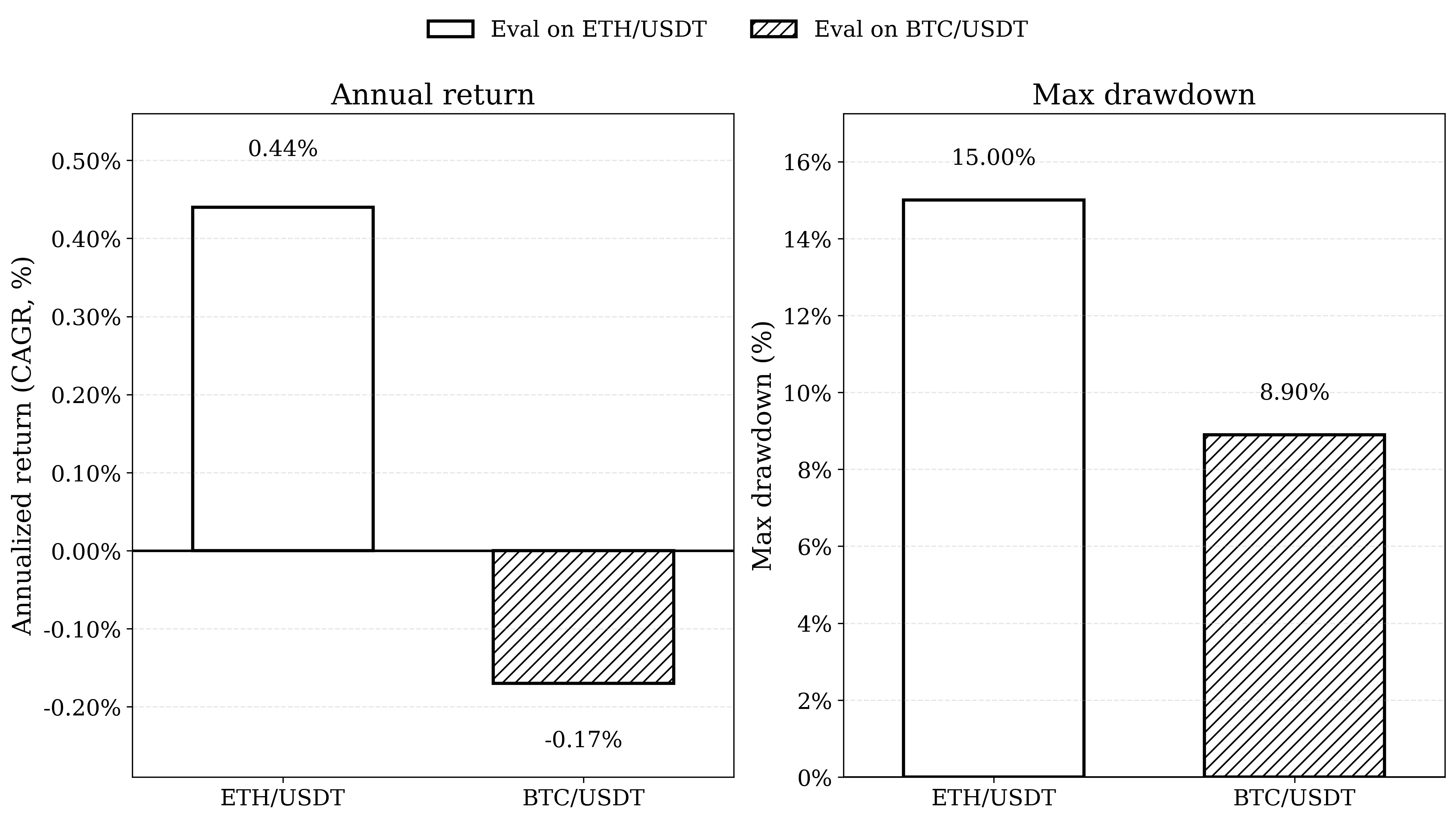}
  \caption{Cross-asset robustness of the ETH-trained STRICT4H aggressive configuration evaluated on ETH/USDT (in-sample) and BTC/USDT (cross-asset) over the overlapping data window.}
  \label{fig:cross_asset}
\end{figure}

Third, to test portability beyond a single contract, an ETH-trained STRICT4H
aggressive configuration is evaluated on BTC/USDT over the
overlapping window. Table~\ref{tab:eth_strict4h_main} reports the key
summary metrics under consistent rigorous trading-cost profiles, and
Figure~\ref{fig:cross_asset} visualizes annual return and maximum drawdown
for the in-sample (ETH) and cross-asset (BTC) evaluations. The exercise is a
diagnostic portability check under identical accounting and execution
semantics.

\begin{table}[htbp]
  \centering
  \footnotesize
  \setlength{\tabcolsep}{3pt}
  \caption{ETH STRICT4H cross-asset performance under consistent realistic trading-cost profiles for ETH and BTC over the overlapping data window (annual return reported as CAGR in \%).}
  \label{tab:eth_strict4h_main}
  \begin{tabular}{@{}>{\raggedright\arraybackslash}p{0.55\columnwidth}ccrrr@{}}
    \toprule
    Strategy & Train & Eval & Ann.\ (\%) & MaxDD & Trades \\
    \midrule
    ETH STRICT4H aggressive v1 & ETH & ETH & 0.44\% & 0.150 & 3{,}100 \\
    ETH STRICT4H aggressive v1 & ETH & BTC & -0.17\% & 0.089 & 3{,}403 \\
    \bottomrule
  \end{tabular}
\end{table}

\subsection{Auxiliary Out-of-Family Check}
To test whether the validation protocol remains informative outside the main
STRICT4H family, the study includes one auxiliary BTC/USDT 4h experiment based
on a long-only volatility-momentum family. The check preserves the same train,
validation, and post-2024 window logic, applies the same strict execution and
cost semantics, and evaluates candidates under the same Stage~II threshold
style used in the BTC case study. For this supplementary check, a
deterministic 120-trial Stage~I search budget is used, and the
top-30 candidates were screened over the same \(3 \times 3\) taker/funding scenario grid.
Under this larger auxiliary search, one candidate passed the pre-registered
Stage~II thresholds.

\begin{table}[htbp]
  \centering
  \small
  \setlength{\tabcolsep}{4pt}
  \caption{\textbf{Auxiliary BTC/USDT 4h volatility-momentum family check under the same strict screening logic.} The one-stage top-1 candidate is reported alongside the best Stage~II candidate under the 120-trial follow-up search. Annual return is reported as CAGR in decimal units.}
  \label{tab:aux_family_check}
  \begin{tabular}{@{}llrrr@{}}
    \toprule
    Candidate & Window & Ann. return & Monthly geom & MaxDD \\
    \midrule
    One-stage top-1 & Training & 0.3046 & 0.0225 & 0.193 \\
    One-stage top-1 & Validation & 0.0000 & 0.0000 & 1.000 \\
    One-stage top-1 & Post-2024 & 0.1469 & 0.0112 & 0.154 \\
    \midrule
    Best Stage~II candidate & Training & 0.0197 & 0.0016 & 0.102 \\
    Best Stage~II candidate & Validation & 0.0772 & 0.0062 & 0.061 \\
    Best Stage~II candidate & Post-2024 & -0.0052 & -0.0004 & 0.100 \\
    \bottomrule
  \end{tabular}
\end{table}

The auxiliary results clarify protocol behavior outside the main family in two
ways. First, the one-stage top-1 candidate looked attractive in the
training window but collapsed on validation, showing the same overfitting risk
that motivated the double-screening protocol in the main BTC study. Second,
one candidate did pass the Stage~II acceptance thresholds under the cost-stress
grid, but the resulting configuration remained modest: its validation
mean monthly geometric return was 0.0062 and its post-2024 annual return was
slightly negative. This auxiliary check gives bounded empirical evidence that
the protocol can be applied beyond the main signal family.

To check whether the evaluator can accommodate optimizer choices beyond TPE, a
bounded auxiliary optimizer check was also run on the same auxiliary volatility-momentum
family. Random search, differential evolution (DE), a simple genetic algorithm
(GA), and Gaussian-process Bayesian optimization (GP-BO) were each given 18
strict train-window evaluations and were then evaluated on the validation and
post-2024 windows under the same cost profile.
Table~\ref{tab:aux_optimizer_check} shows that alternative black-box optimizers
can be routed through the same STRICT4H evaluator. It also shows that several
methods find train-window candidates whose validation behavior is weak or
degenerate, which is precisely the kind of fragility the screening layer is
designed to expose.
\begin{table}[htbp]
  \centering
  \small
  \setlength{\tabcolsep}{4pt}
  \caption{\textbf{Bounded auxiliary optimizer check on the BTC/USDT 4h volatility-momentum family.} All methods use the same STRICT4H train-window objective and the same low evaluation budget. The table reports bounded substitutability evidence.}
  \label{tab:aux_optimizer_check}
  \begin{tabular}{@{}lrrrrrr@{}}
    \toprule
    Method & Eval. & Best train & Val ann. & Val MaxDD & Post ann. & Post MaxDD \\
    \midrule
    Random & 18 & 0.1832 & 0.0241 & 0.0243 & -0.0003 & 0.0944 \\
    DE & 18 & 0.0992 & -0.0006 & 0.1460 & 0.0688 & 0.0977 \\
    GA & 18 & 0.1579 & 0.0000 & 1.0000 & 0.1890 & 0.2195 \\
    GP-BO & 18 & 0.1352 & 0.0000 & 1.0000 & 0.2002 & 0.1904 \\
    \bottomrule
  \end{tabular}
\end{table}

\subsection{Main-Family Optimizer Comparison and Component Diagnostics}
The auxiliary optimizer check is complemented by a same-budget benchmark on the
main BTC/USDT STRICT4H family. TPE, random search, DE, a
simple GA, and GP-based Bayesian optimization were routed through the same
shared parameter space, objective, cost profile, and 60-evaluation budget.
Table~\ref{tab:main_optimizer_benchmark} reports sampler-comparison evidence:
in this bounded run, GP-BO produced the strongest training candidate and the best
post-2024 monthly geometric return among the five methods, while all methods'
best training candidates were weak on the 2021--2022 validation window. This
is consistent with the paper's core claim that optimizer output must be passed
through strict screening rather than accepted at face value.
\begin{table*}[htbp]
  \centering
  \scriptsize
  \setlength{\tabcolsep}{3pt}
  \caption{\textbf{Main-family same-budget optimizer benchmark on the BTC/USDT STRICT4H training window.} All optimizers use the same shared main-family parameter space, objective, cost profile, and evaluation budget. Validation and post-2024 metrics evaluate the best training-window candidate found by each optimizer. Mon. denotes monthly geometric return, DD denotes maximum drawdown, and tr. denotes trades. The benchmark is interpreted as bounded evaluator-comparison evidence.}
  \label{tab:main_optimizer_benchmark}
  \begin{tabular}{@{}lrrrrrrrr@{}}
    \toprule
    Method & Eval. & Invalid \% & Train obj. & Val mon. & Val DD & Post mon. & Post DD & Post tr. \\
    \midrule
    TPE & 60 & 36.7 & 0.1870 & 0.0000 & 1.000 & 0.0159 & 0.545 & 437 \\
    Random & 60 & 66.7 & 0.0748 & 0.0000 & 1.000 & 0.0166 & 0.385 & 364 \\
    DE & 60 & 63.3 & 0.0450 & 0.0000 & 1.000 & -0.0151 & 0.427 & 203 \\
    GA & 60 & 28.3 & 0.0498 & 0.0000 & 1.000 & -0.0003 & 0.356 & 2204 \\
    GP-BO & 60 & 45.0 & 0.2184 & 0.0000 & 1.000 & 0.0400 & 0.291 & 867 \\
    \bottomrule
  \end{tabular}
\end{table*}

Selected decision and execution modules were then disabled on the best
main-family candidate returned by each optimizer. Table~\ref{tab:module_ablation}
reports the post-2024 effect on the GP-BO candidate and the median effect
across the five optimizer-selected candidates. Disabling the ADX/regime gate
substantially increased drawdown and turnover; disabling confirmation and
turnover controls increased switching; disabling signal pre-averaging reduced
post-2024 monthly return in the median candidate. Funding-rule removal was
near-neutral in the median post-2024 comparison, although it changed the
individual GP-BO candidate's return and drawdown. The
module ablation is therefore interpreted as component-level diagnostic evidence.
\begin{table*}[htbp]
  \centering
  \scriptsize
  \setlength{\tabcolsep}{3pt}
  \caption{\textbf{Module ablation on the main-family benchmark candidates.} The table shows the post-2024 effect on the GP-BO best training candidate and, for ablated variants, the median post-2024 change across the five optimizer-selected candidates. Mon. denotes monthly geometric return, DD denotes maximum drawdown, tr. denotes trades, and sw. denotes switch density. Positive \(\Delta\)DD means larger drawdown. The experiment is a bounded component diagnostic, not a universal component ranking.}
  \label{tab:module_ablation}
  \begin{tabular}{@{}lrrrrrr@{}}
    \toprule
    Variant & GP mon. & GP DD & GP tr. & Med. \(\Delta\)mon. & Med. \(\Delta\)DD & Med. \(\Delta\)sw. \\
    \midrule
    Full & 0.0400 & 0.291 & 867 & -- & -- & -- \\
    NoFunding & 0.0490 & 0.231 & 867 & +0.0000 & +0.000 & -0.0005 \\
    NoPreAvg & 0.0129 & 0.181 & 660 & -0.0197 & -0.161 & -0.0123 \\
    NoRegime & 0.0256 & 0.848 & 3643 & -0.0045 & +0.443 & +0.1283 \\
    NoConfirmTurn & 0.0413 & 0.303 & 3567 & +0.0025 & +0.013 & +0.1089 \\
    NoBoostStruct & 0.0400 & 0.291 & 867 & +0.0000 & +0.000 & +0.0000 \\
    \bottomrule
  \end{tabular}
\end{table*}

Finally, alternative Stage~II policies were applied to the same valid top-20 main-family
trial pool. Table~\ref{tab:stage2_policy_ablation} varies the rolling
step size, threshold strictness, and cost-grid inclusion while keeping the
candidate pool fixed. No candidate passed even under the relaxed threshold or
the no-cost-grid variant, indicating that the same-budget optimizer candidates
remain weak selections under the paper's stability policy.
\begin{table*}[htbp]
  \centering
  \scriptsize
  \setlength{\tabcolsep}{3pt}
  \caption{\textbf{Stage-II rolling-step and screening-threshold ablation.} Each policy screens the same valid main-family trial pool over the validation window. The base policy uses 2,000-bar rolling windows stepped by 500 bars and the ex-ante thresholds in Table~\ref{tab:stable_candidate_rules}.}
  \label{tab:stage2_policy_ablation}
  \begin{tabular}{@{}lrrrrrrrr@{}}
    \toprule
    Policy & Step & Scen. & Pass & Mean thr. & Min thr. & MaxDD cap & Switch cap & Top pass monthly \\
    \midrule
    Base & 500 & 9 & 0 & 0.0050 & 0.0000 & 0.30 & 0.12 & -- \\
    Step 250 & 250 & 9 & 0 & 0.0050 & 0.0000 & 0.30 & 0.12 & -- \\
    Step 1000 & 1000 & 9 & 0 & 0.0050 & 0.0000 & 0.30 & 0.12 & -- \\
    Relaxed & 500 & 9 & 0 & 0.0030 & -0.0020 & 0.40 & 0.18 & -- \\
    Strict & 500 & 9 & 0 & 0.0075 & 0.0020 & 0.25 & 0.08 & -- \\
    No cost grid & 500 & 1 & 0 & 0.0050 & 0.0000 & 0.30 & 0.12 & -- \\
    \bottomrule
  \end{tabular}
\end{table*}

\section{Discussion}\label{sec:discussion}

This section interprets the empirical evidence as decision-support evidence
for execution-aware configuration selection. The discussion focuses on what
the experiments imply about execution realism, configuration fragility,
residual overfitting risk, and the practical boundaries of deployment-oriented
decision support in perpetual futures.

\paragraph{Interpretation of the empirical contribution}
These experiments should therefore be read as evidence about the
behavior of an execution-aware validation protocol. The contribution is not a
new alpha engine; it is a disciplined way to show when apparent profitability
depends on optimistic execution assumptions, fragile window selection, or
insufficiently explicit governance of configuration choice.

\paragraph{Scope boundaries and limitations}
This manuscript emphasizes five scope limits. First, the optimizer evidence is bounded:
although the paper reports a same-budget comparison across TPE, random
search, DE, GA, and GP-based Bayesian optimization, the result is a bounded
single-study diagnostic rather than a universal optimizer ranking. Second,
although the manuscript documents an explicit OHLCV and
funding-construction protocol, exchange-specific data-construction choices can
still affect perpetual-futures studies and remain a valid target for future
data-pipeline checks. Third, the trading-cost model is linear in turnover;
nonlinear market impact and venue-specific liquidity constraints are left to
future execution models. Fourth, the core evidence remains restricted to a bounded strategy
family and a small set of liquid perpetual contracts; the cross-asset results,
module checks, and one auxiliary out-of-family check are portability
and fragility diagnostics for the protocol. Fifth, although the implementation includes third-party replay checks, a local
Freqtrade adapter backtest, deployment-oriented parity checks, and guard-based
supervision, broader third-party integrations and production exchange
deployment remain future work.

\paragraph{Protocol portability beyond one signal family}
Architecturally, AutoQuant is signal-modular: the strict engine consumes a
parameterized signal-and-risk specification, while the traceability protocol operates
at the level of execution semantics, cost realism, screening policy, and
artifact traceability. For this reason, the contribution is not limited to one
specific momentum rule. At the implementation level, the codebase exposes
multiple signal blocks, including momentum, mean-reversion, breakout,
volatility-momentum, structural, and factor-style components, under the same
strict execution engine. Repository tests confirm the narrower point that the
same runtime can ingest
different parameter branches, including momentum, mean-reversion, and blended
multi-leg specifications, without changing the underlying execution semantics.
To probe this point empirically, a bounded BTC/USDT 4h auxiliary
check using a long-only volatility-momentum family under the same window logic
and Stage~II screening rules was also run; the one-stage top-1 candidate collapsed on
validation and one candidate passed the Stage~II acceptance thresholds. This
shows that the protocol can expose fragility outside the main signal family
while keeping strategy-family generality as a future empirical question.
Implementation-level checks therefore support protocol portability within the
reported scope.

\paragraph{Residual selection-bias diagnostics}
Even under strict execution semantics and explicit cost modeling, repeated
search over candidate configurations leaves non-trivial multiple-testing risk.
The manuscript therefore reports two selection-bias diagnostics from the
reproducible BTC/USDT anchor snapshot. First, following the spirit of
\citep{bailey2014deflated}, a DSR-style check based on the long-window Sharpe
and an effective trial-count range \(N_{\mathrm{total}} \in [360,3240]\) is
consistent with non-negligible search risk once repeated trials are taken into
account. Second, a CSCV/PBO diagnostic \citep{bailey2017pbo} computed on the
top-40 Stage~I candidate pool yields \(\mathrm{PBO}=0.586\) under an 8-segment
design with 70 combinatorial splits. These diagnostics reinforce the
manuscript's emphasis on traceability and bias control under explicit
execution rules.

\paragraph{Optimizer scope}
Although TPE is justified as a practical sampler for conditional mixed search
spaces, the optimizer evidence remains bounded. In addition to the original
TPE-versus-random comparison, this paper includes a main-family same-budget
comparison with DE, a simple GA, and GP-based Bayesian optimization. This
broader check shows heterogeneous optimizer behavior: GP-BO is strongest in
the reported training-window run, while the resulting candidates do not satisfy
the validation screening policies. Accordingly, the contribution lies in the
execution- and traceability-preserving configuration-selection pipeline, not
in an optimizer-ranking result.

\paragraph{Data-construction disclosure}
Section~\ref{data-and-market-setting} documents the core OHLCV
construction protocol, including UTC bar-open indexing, sanity filtering,
de-duplication, and continuity validation, and points to the scripts used
to rebuild long-window 4-hour candles and funding baselines. Nevertheless,
several exchange-specific data-construction choices can still matter in
perpetual markets, including whether a venue's OHLCV reflects last-trade,
mark, or index pricing, cross-venue aggregation rules, and the handling
of contract-specification changes. These details remain a natural target
for future data-construction work.

\paragraph{Linear cost model and position scaling}
The cost profile used in the STRICT4H experiments assumes an initial
capital of 10,000 USDT, a maximum leverage of \(5\times\), and a notional
cap of 50,000 USDT. The reported annualized returns and Sharpe ratios
should therefore be interpreted as describing small- to mid-sized
positions in a frictional but non-impact world. The model does not include
non-linear market impact or exchange-specific liquidity constraints, so
the same percentage returns should not be extrapolated to institutional asset
levels. Consistent with this view, very large
final-equity figures that arise in long synthetic compounding runs are
treated as accounting artifacts and deliberately omitted from the main
tables. More broadly, the cost model is linear in turnover and does not
capture non-linear impact; incorporating impact-aware execution models is
outside the scope of the current manuscript.

\paragraph{Alpha logic constraints}
While the framework is signal-modular at the architectural level,
integrating highly complex, non-parametric machine-learning models would
require additional engineering. The current auto-tuning pipeline is not
designed to manage the separate training, validation, and GPU resource
management loops required by such models, and the empirical evidence in
this paper remains centered on a STRICT4H momentum-style family on
BTC/USDT, ETH/USDT, SOL/USDT, and AVAX/USDT, plus one auxiliary BTC/USDT 4h
volatility-momentum family check that yielded one Stage~II-stable candidate in
the larger follow-up search.
Relatedly, automated
\emph{candidate generation} such as factor-library composition, template
expansion, or symbolic regression is outside the scope of this manuscript.
Such generators can still be evaluated as upstream modules under the same
strict \(t\!+\!1\) semantics, cost modeling, and multi-window screening
used here.

\section{Conclusion and Future Work}\label{sec:conclusion}

This paper introduced a systematic framework for the auto-tuning and
multi-stage validation of quantitative strategies for cryptocurrency perpetual
futures. Using BTC/USDT as the anchor case and replications on
ETH/USDT, SOL/USDT, and AVAX/USDT, the analysis showed that backtests which neglect
realistic costs or permit subtle look-ahead effects can materially overstate
strategy quality.

The primary contribution is a structured two-stage selection pipeline
complemented by a post-deployment monitoring layer. Concretely, the
framework combines Bayesian optimization under realistic cost models, a
double-screening protocol using held-out and cost-stress scenarios, and an
traceability/guard overlay. The result is an auditable decision-support and
governance protocol for configuration selection under strict execution
semantics.

Across the four contracts, realistic execution and costs materially change
candidate rankings and reveal degradation on held-out windows. The non-BTC
replications provide pipeline-level portability checks under identical strict
semantics, and an auxiliary out-of-family BTC/USDT check further shows that
the same protocol can reject unstable candidates outside the main
signal family. These supplements broaden the evidence base and clarify the empirical
scope of the framework.
By enforcing full-chain accounting invariants and employing a guard
mechanism, the framework provides a practical research-to-deployment pattern
for this asset class. That is presented as a design objective supported by
offline and replay-style consistency evidence.

From an expert-systems and decision-support perspective, the practical
implication is that quantitative evidence in 24/7 derivatives markets should
be delivered as an \emph{auditable artifact}: a complete specification of
timestamp conventions, execution semantics, cost and funding alignment, and
window-reuse rules, together with deterministic scripts that regenerate the
reported tables and figures. Appendix~A therefore includes both
full-chain consistency checks and minimal external semantics
replays.

\noindent\textbf{Managerial implications.}
For practitioners building algorithmic trading or risk-control pipelines in
24/7 derivatives markets, the evidence supports four governance practices:
treat execution semantics and funding alignment as non-negotiable policies
rather than optional backtest settings; pre-register window semantics and
selection rules to avoid hidden reuse of validation segments; maintain
deterministic per-run artifacts for independent inspection; and deploy
lightweight guard overlays that monitor invariant violations and reconcile
offline and live-style replays.

Finally, the numeric results reported in the tables and figures are
accompanied by the unified codebase, anonymized replication materials, and
derived summary artifacts described in Appendix~A, subject to the data-access
and storage constraints discussed there. A central limitation remains that the present
cost model is linear in turnover. Future work should incorporate nonlinear
market impact, liquidity-dependent slippage, queue-position effects, and
venue-specific execution constraints before drawing capacity or institutional
deployment conclusions.

\appendix
\section{Reproducibility, Code, Data, and AI Tools}\label{app:repro}

This paper places a strong emphasis on reproducibility and transparency. The
experiments reported in this paper were prepared within a unified codebase
that implements the backtest engine, Bayesian optimization pipeline,
double-screening protocol, and live-guard logic described in the main text.
For this manuscript, an anonymized replication package is maintained, built from
the same codebase, that provides the core 4-hour datasets and/or scripts to
regenerate them, together with configuration files, derived summary artifacts,
and helper scripts needed to reproduce and check the reported backtests, validation
procedures, and tables. This package can be shared with qualified researchers
on reasonable request, subject to exchange
data-licensing and storage constraints. The numeric results shown in the
manuscript are reported from simulation outputs and derived machine-readable
summaries so that the paper remains aligned with the underlying experiment
logs; where only summary-level outputs are retained in the submission snapshot,
the associated tables are treated as bounded empirical summaries rather than as
fresh full reruns.

For the core tables and figures, the replication package includes deterministic
scripts and summary artifacts that support regeneration or reconciliation of
the machine-readable outputs used to produce the reported results, subject to
the availability of the required market-data snapshots. A brief README lists
the commands, expected output filenames, input-data requirements, and runtime
notes. The package also includes a helper script that regenerates the figures
used in this \LaTeX{} submission.

The replication snapshot includes selected derived machine-readable artifacts
and supporting reproducibility outputs that are used to reconcile the manuscript tables
and figures. End-to-end regeneration scripts for the data-dependent tables,
together with optional third-party semantics replays, are provided in the
companion replication package; because full-chain logs can be large, this
snapshot keeps selected derived artifacts used in the paper rather than every
intermediate output. The auxiliary out-of-family check and semantic/funding
checks are directly machine-readable. The one-stage/two-stage validation rows and the
cost-ablation ladder also have table consistency checks against
companion summary artifacts. Several other large historical
result tables remain traceable through companion summary artifacts rather
than complete local rerun logs.

\noindent\textbf{Execution-semantics reproducibility checks.}
The repository also includes regression tests and consistency checks for the
strict \(t\!+\!1\) execution convention, cost accounting, funding visibility,
and selected carried-forward tables. These checks keep the manuscript's
empirical claims aligned with the submitted tables and reproducibility
artifacts. Same-bar \(t\!+\!0\) sensitivity results are reported below as
execution-semantics diagnostics rather than selected-performance evidence.

\begin{table}[htbp]
  \centering
  \footnotesize
  \setlength{\tabcolsep}{4pt}
  \caption{\textbf{Minimal reproducibility artifacts exported by AutoQuant.} Each run exports machine-readable files that support table/figure regeneration or reconciliation and decompose net returns into raw-return and cost components. Exact filenames and schemas are documented in the replication README.}
  \label{tab:repro_artifacts}
  \begin{tabular}{@{}p{0.28\linewidth}p{0.30\linewidth}p{0.38\linewidth}@{}}
    \toprule
    Artifact (format) & Contents (examples) & Reproducibility purpose \\
    \midrule
    Configuration (\texttt{json}) & \(\theta\), cost profile, window spec, seeds & Reproducible run specification and provenance \\
    Per-bar ledger (\texttt{csv}) & timestamp, signal \(S_t\), exposure \(\pi_t\), \(r^{\mathrm{mkt}}_t\), \(r^{\mathrm{raw}}_t\), \(C_{\mathrm{fee},t}\), \(C_{\mathrm{slip},t}\), \(C_{\mathrm{fund},t}\), \(r^{\mathrm{net}}_t\) & Full reconstruction of metrics and cost decomposition \\
    Robust summary (\texttt{csv}) & window/scenario aggregates (monthly geom, MaxDD, turnover) & Stage~II screening inputs and sensitivity reporting \\
    Invariant check (\texttt{json/csv}) & max \(|\Delta S|\), max \(|\Delta\pi|\), component diffs & Backtest-to-replay consistency checks \\
    \bottomrule
  \end{tabular}
\end{table}

\noindent\textbf{Execution-semantics diagnostic.}
As an additional minimal check reproducible from the replication snapshot, the
appendix reports how performance changes when the same 4-hour strategy parameters are
evaluated under strict \(t\!+\!1\) execution versus a deliberately naive
same-bar (\(t\!+\!0\)) convention. This semantic sensitivity check is separate
from parameter selection. Values in Table~\ref{tab:semantics_sanity} are generated
from the replication artifacts described above.

\begin{table*}[htbp]
  \centering
  \footnotesize
  \setlength{\tabcolsep}{5pt}
  \caption{\textbf{Execution-semantics diagnostic (BTC/USDT 4h; 2022-01-01--2023-01-01).} Fixed parameters; only the execution delay differs: strict \(t\!+\!1\) versus same-bar \(t\!+\!0\). Annual return is CAGR in decimal units; monthly geometric mean is in decimal units; MaxDD is a fraction in \([0,1]\).}
  \label{tab:semantics_sanity}
  \begin{tabular}{@{}lrrrrr@{}}
    \toprule
    Semantics & Ann. CAGR & Monthly geom & Sharpe & MaxDD & Trades \\
    \midrule
    STRICT (\(t\!+\!1\)) & -0.221 & -0.021 & -2.413 & 0.221 & 30 \\
    NAIVE (\(t\!+\!0\)) & -0.264 & -0.025 & -2.998 & 0.264 & 28 \\
    \bottomrule
  \end{tabular}
\end{table*}

\noindent\textbf{Batch semantics uplift (Stage~I top-\(N\)).}
To avoid over-interpreting a single parameter setting, the appendix also reports a batch
summary over the top-\(N\) Stage~I trials. Each trial is re-evaluated on the
same hold-out window under STRICT \(t\!+\!1\) versus NAIVE \(t\!+\!0\), with
all other settings fixed. This check asks whether same-bar execution
\emph{systematically} inflates performance in the snapshot. For BTC/USDT on
2022-01-01--2023-01-01 using the top \(N=30\) Stage~I trials, the median
uplift (NAIVE minus STRICT) in annualized return is negative and the fraction
of trials with positive uplift is below 50\%, indicating that naive same-bar
execution is not a universal performance booster in this setting.

\begin{table}[htbp]
  \centering
  \small
  \setlength{\tabcolsep}{5pt}
  \caption{\textbf{Batch execution-semantics uplift (BTC/USDT 4h; 2022-01-01--2023-01-01; top \(N=30\) Stage~I trials).} Uplift is NAIVE(\(t\!+\!0\)) minus STRICT(\(t\!+\!1\)) in annualized return (CAGR, decimal units).}
  \label{tab:semantics_uplift_batch}
  \begin{tabular}{@{}rrrrr@{}}
    \toprule
    \(N\) & Median uplift & 25th pct & 75th pct & \% uplift \(>0\) \\
    \midrule
    30 & -0.018 & -0.361 & 0.033 & 0.367 \\
    \bottomrule
  \end{tabular}
\end{table}

\noindent\textbf{Third-party semantics replay (Backtrader).}
As a minimal external check of the strict close-to-close \(t\!+\!1\)
raw-return accounting, the same directional signal series is replayed in
Backtrader under cheat-on-close execution, and the resulting per-bar
raw-return series is compared to the vectorized STRICT4H construction. On BTC/USDT 4h over
2022-01-01--2023-01-01, the maximum absolute per-bar difference is on the
order of \(10^{-6}\) and the correlation is effectively 1.0, supporting the
claim that the strict raw-return semantics are not an artifact of a single
bespoke implementation.

\begin{table}[htbp]
  \centering
  \small
  \setlength{\tabcolsep}{5pt}
  \caption{\textbf{Backtrader raw-return replay check (BTC/USDT 4h; 2022-01-01--2023-01-01).} Raw returns only; costs and funding are disabled for this semantics-only check.}
  \label{tab:backtrader_semantics}
  \begin{tabular}{@{}rrrr@{}}
    \toprule
    Bars & Max \( |\Delta| \) & RMSE & Corr. \\
    \midrule
    2190 & \(2.75\times 10^{-6}\) & \(2.94\times 10^{-7}\) & 1.000 \\
    \bottomrule
  \end{tabular}
\end{table}

\noindent\textbf{Third-party framework replay benchmark.}
For external framework comparison, a Backtrader adapter replay was also run on
the main-family GP-BO candidate. The adapter uses
Backtrader's event loop to replay the executed STRICT4H exposure series and
then applies the same fee, slippage, and funding ledger used by AutoQuant.
On the post-2024 BTC/USDT window, the maximum absolute per-bar net-return
difference was \(5.44\times10^{-5}\), with RMSE \(4.68\times10^{-6}\) and a
return correlation effectively equal to one. Across the top-5 main-family
candidates and five rolling validation windows, the rank
correlation between AutoQuant and Backtrader replay metrics was 1.000, with
maximum monthly-metric difference \(3.14\times10^{-4}\). This supports
semantic portability when a custom adapter and explicit ledger are supplied.
Freqtrade was also evaluated with its local futures backtester through a
programmatic signal adapter derived from the same GP-BO exposure path. The
Freqtrade run completed with 65 trades from 265 exposure changes, total trade
PnL of 137.87\%, and maximum account drawdown of 16.19\%. This is reported as a
framework-adapter benchmark with explicit adapter semantics.
\begin{table*}[htbp]
  \centering
  \scriptsize
  \setlength{\tabcolsep}{3pt}
  \caption{\textbf{Third-party framework replay benchmark.} Backtrader is evaluated through an event-loop adapter over the same STRICT4H executed exposure series and cost/funding profile. Freqtrade is evaluated with its local futures backtester through a programmatic signal adapter derived from the AutoQuant GP-BO exposure path. The table reports an offline framework-replay comparison.}
  \label{tab:framework_replay_benchmark}
  \begin{tabular}{@{}p{0.18\linewidth}p{0.16\linewidth}p{0.17\linewidth}p{0.16\linewidth}p{0.25\linewidth}@{}}
    \toprule
    Path & Run status & Backtest deviation & Stability check & Scope note \\
    \midrule
    AutoQuant STRICT4H & completed & reference & reference top-5 pool & native per-bar signal, exposure, costs, funding, and net-return ledger \\
    Backtrader adapter & completed & max 5.44e-05; RMSE 4.68e-06 & rank rho 1.000; max monthly delta 3.14e-04 & custom exposure replay with explicit cost/funding ledger \\
    Freqtrade adapter & completed & trade PnL 137.87\%; DD 16.19\% & 65 trades from 265 exposure changes & local futures backtester with signal adapter \\
    \bottomrule
  \end{tabular}
\end{table*}

\noindent\textbf{Runtime parity check.}
Beyond the synthetic and accounting-style checks above, the repository also
includes an automated strict-\(t\!+\!1\) parity script that compares the
rigorous backtest path with the live-style inference wrapper on the same data
slice and parameter snapshot. This check was run on three independent
SOL/USDT 1h parameter snapshots over 2{,}000 bars
each, with exact agreement in the exported signal and exposure series
(\texttt{direction\_match\_ratio}=1.0, maximum absolute signal difference \(=0\),
maximum absolute exposure difference \(=0\)). This provides code-path
consistency evidence for the offline-to-live bridge.

The historical BTC/USDT, ETH/USDT, SOL/USDT, and AVAX/USDT perpetual data used
in this study were obtained from major centralized exchanges, primarily
Binance, Bybit, and OKX. Due to exchange licensing and storage constraints, the replication package does
not redistribute raw tick-level data. The replication materials therefore
include scripts and configuration templates that allow qualified researchers to
re-download data from original sources, subject to exchange terms, and to
regenerate the cleaned 4-hour OHLCV and funding-rate series used in the
experiments. Processed summary statistics and anonymized walk-forward tables
can be shared with qualified researchers on reasonable request, subject to
data-licensing constraints.

\noindent\textbf{High-frequency guard experiments.}
As part of the broader project, a series of high-frequency ATR-guard
experiments on 1m--60m resampled BTC/USDT data were also conducted. These
auxiliary experiments are not part of the main empirical results reported in this paper.
Empirically, none of the tested configurations delivered a clear improvement in
risk-adjusted performance once realistic fees and slippage were taken into
account; most settings either left performance essentially unchanged or
degraded it by introducing noise-induced turnover. These results are therefore interpreted
as context-specific evidence that, in the high-friction cost
regime considered here, a 4-hour decision horizon is more suitable than naive
1m--15m overlays, while recognizing that more sophisticated high-frequency
designs would require separate study.

\section*{Declaration of generative AI and AI-assisted technologies}
During the preparation of this manuscript, the author used OpenAI Codex CLI in a
limited assistive capacity for English copy-editing and tightening of
non-technical prose, and for drafting reproducibility and table/figure
traceability checklists. After using this tool, the author reviewed and edited
the content as needed and takes full responsibility for the content of the
published article. All experimental results, numerical findings, and
conclusions reported in this manuscript were generated and verified solely by
the author using the described codebase and exported artifacts. This AI tool
had no role in producing data, selecting configurations, or computing metrics.

\bibliographystyle{elsarticle-num}
\bibliography{sn-bibliography}

\end{document}